\pdfoutput=1
\documentclass[11pt,twoside,a4paper,cmspaper,final,collab]{cms-tdr}
\def\svnVersion{6507df4}\def\svnDate{2023/05/18}\def\cmsCernNoTag{CERN-EP-2022-031}\def\cmsCernDate{\today}\def\cmsMessage{Published in Physics Letters B as \href{http://dx.doi.org/10.1016/j.physletb.2022.137460}{\doi{10.1016/j.physletb.2022.137460}.}}
\begin{document}\cmsNoteHeader{SUS-21-002}

\newlength\cmsTabSkip\setlength{\cmsTabSkip}{1ex}
\ifthenelse{\boolean{cms@external}}{\providecommand{\cmsLeft}{upper\xspace}}{\providecommand{\cmsLeft}{left\xspace}}
\ifthenelse{\boolean{cms@external}}{\providecommand{\cmsRight}{lower\xspace}}{\providecommand{\cmsRight}{right\xspace}}

\newcommand{\oneell}{\ensuremath{1\Pell}\xspace}
\newcommand{\zeroell}{\ensuremath{0\Pell}\xspace}
\newcommand{\oneres}{1-res\xspace}
\newcommand{\zerores}{0-res\xspace}

\newcommand{\mJ}{\ensuremath{m_{\text{J}}}\xspace}
\newcommand{\nb}{\ensuremath{n_{\PQb}}\xspace}
\newcommand{\muR}{\ensuremath{\mu_\mathrm{R}}\xspace}
\newcommand{\muF}{\ensuremath{\mu_\mathrm{F}}\xspace}

\newcommand{\Wqqprime}{\ensuremath{\PW(\PQq\HepParticle{\PAQq}{}{\prime})}\xspace}
\newcommand{\zjets}{\ensuremath{\PZ+\text{jets}}\xspace}
\newcommand{\wjets}{\ensuremath{\PW+\text{jets}}\xspace}
\newcommand{\vjets}{\ensuremath{\PW/\PZ+\text{jets}}\xspace}
\newcommand{\ttjets}{\ensuremath{\PQt\PAQt+\text{jets}}\xspace}
\renewcommand{\bbbar}{\ensuremath{\PQb\PAQb}\xspace}
\newcommand{\ttH}{\ensuremath{\PQt\PAQt\PH}\xspace}
\newcommand{\WH}{\ensuremath{\PW\PH}\xspace}
\newcommand{\WW}{\ensuremath{\PW\PW}\xspace}
\newcommand{\WZ}{\ensuremath{\PW\PZ}\xspace}
\newcommand{\ZH}{\ensuremath{\PZ\PH}\xspace}

\newcommand{\fullLumi}{137\fbinv}
\newcommand{\thirdchi}{\PSGczDT}
\newcommand{\secondchi}{\PSGczDt}
\newcommand{\firstchi}{\PSGczDo}
\newcommand{\firstcharg}{\PSGcpmDo}
\newcommand{\chargneut}{\ensuremath{\firstcharg\secondchi}\xspace}
\newcommand{\chargneutthree}{\ensuremath{\firstcharg\thirdchi}\xspace}
\newcommand{\neuttwoneutthree}{\ensuremath{\secondchi\thirdchi}\xspace}
\newcommand{\firstchargpair}{\ensuremath{\firstcharg\HepParticle{\PSGc}{1}{\mp}}\xspace}
\newcommand{\charginoi}{\ensuremath{\HepParticle{\PSGc}{\mathrm{i}}{\pm}}\xspace}
\newcommand{\neutralinoi}{\ensuremath{\HepParticle{\PSGc}{\mathrm{i}}{0}}\xspace}
 
\cmsNoteHeader{SUS-21-002}
\title{Search for electroweak production of charginos and neutralinos at \texorpdfstring{$\sqrt{s}=13\TeV$}{sqrt(s)=13 TeV} in final states containing hadronic decays of \texorpdfstring{\WW, \WZ, or \WH}{WW, WZ, or WH} and missing transverse momentum}

\date{\today}

\abstract{
This Letter presents a search for direct production of charginos and neutralinos via electroweak interactions. The results are based on data from proton-proton collisions at a center-of-mass energy of 13\TeV collected with the CMS detector at the LHC, corresponding to an integrated luminosity of \fullLumi. The search considers final states with large missing transverse momentum and pairs of hadronically decaying bosons \WW, \WZ, and \WH, where \PH is the Higgs boson. These bosons are identified using novel algorithms. No significant excess of events is observed relative to the expectations from the standard model. Limits at the 95\% confidence level are placed on the cross section for production of mass-degenerate wino-like supersymmetric particles \firstcharg and \secondchi, and mass-degenerate higgsino-like supersymmetric particles \firstcharg, \secondchi, and \thirdchi. In the limit of a nearly-massless lightest supersymmetric particle \firstchi, wino-like particles with masses up to 870 and 960\GeV are excluded in the cases of $\secondchi\to\PZ\firstchi$ and $\secondchi\to\PH\firstchi$, respectively, and higgsino-like particles are excluded between 300 and 650\GeV.
}

\hypersetup{
pdfauthor={CMS Collaboration},
pdftitle={Search for electroweak production of charginos and neutralinos at sqrt(s)=13 TeV in final states containing hadronic decays of WW, WZ, or WH and missing transverse momentum},
pdfsubject={CMS},
pdfkeywords={CMS, gaugino, electroweakino, supersymmetry}}

\maketitle

\section{Introduction}
\label{sec:intro}

Supersymmetry (SUSY)~\cite{Ramond:1971gb,Golfand:1971iw,Neveu:1971rx,Volkov:1972jx,Wess:1973kz,Wess:1974tw,Fayet:1974pd,Fayet:1976cr,Nilles:1983ge,Martin:1997ns}
proposes to extend the standard model (SM) of particle physics by the addition of a new symmetry.
This new symmetry requires that, for each boson (fermion) in the SM, there is also a fermionic (bosonic) superpartner, also called a ``sparticle''.
SUSY can naturally predict cancellations of large radiative corrections to the Higgs boson (\PH) mass if some sparticles are not too heavy~\cite{Papucci:2011wy}.
Recent work~\cite{Baer:2015tva,PhysRevD.95.055012} suggests that strongly interacting sparticles significantly heavier than current bounds~\cite{Aad:2019pfy,
Aad:2020sgw,Aad:2020aze,Aad:2020aob,Aad:2021zyy,Aad:2021jmg,Aad:2021hjy,Aad:2021oos,
Sirunyan:2019ctn,Sirunyan:2019xwh,CMS:2019tlp,Sirunyan:2019glc,Sirunyan:2020tyy,Sirunyan:2021mrs,CMS:2021eha}
are consistent with a natural theory~\cite{Papucci:2011wy}.
Contrastingly, charginos and neutralinos are still expected to be lighter than
${\sim}1\TeV$, with the lightest having a mass close to that of the Higgs boson.
This suggests that the hunt for \TeVns{}-scale charginos and neutralinos is the next proving ground for natural SUSY.

In this Letter, we present a search for direct electroweak production of sparticles at the CERN LHC.
The superpartners of the gauge bosons of the unbroken $\mathrm{SU}(2)$ and $\mathrm{U}(1)$ symmetries and the superpartners of the Higgs bosons (the winos, bino, and higgsinos, respectively)
mix to form two chargino (\charginoi with $\mathrm{i}=1$, 2)
and four neutralino (\neutralinoi with $\mathrm{i}=1$, 2, 3, 4) mass eigenstates, where $\mathrm{i}$ is in order of increasing mass.
In the case where the mass mixing of the charginos and neutralinos is nearly diagonal, each of the mass eigenstates is identified as bino-, wino-, or higgsino-like, depending on the dominant contribution~\cite{Martin:1997ns}.

We search for direct production of wino-like charginos and neutralinos (\firstcharg and \secondchi)
or higgsino-like charginos and neutralinos (\firstcharg, \secondchi, and \thirdchi).
We assume that the superpartners of the SM leptons are much heavier than the charginos and neutralinos.
As such, decays of wino- or higgsino-like charginos and neutralinos proceed through \PW, \PZ, and Higgs bosons.

In this search, we assume that $R$-parity~\cite{Farrar:1978xj} is conserved and that the lightest neutralino is a bino-like \firstchi that is stable and will escape the detector unobserved.
Such a \firstchi would be a viable candidate for weakly-interacting massive particle dark matter.
We explore four signal regions, which target three prominent signatures when the scalar sparticles
are very heavy: \WW, \WZ, or \WH, together with a large transverse momentum imbalance.

At the LHC, several searches for direct chargino-neutralino and chargino-pair production in these channels have
been performed by the ATLAS~\cite{Aaboud:2018jiw,Aaboud:2018sua,Aaboud:2018ngk,Aad:2019vnb,Aad:2019vvf,Aad:2019vvi,Aad:2021ajl,ATLAS:2021yqv,ATLAS:2022zwa}
and CMS~\cite{Sirunyan:2017zss,Sirunyan:2017lae,Sirunyan:2018ubx,Sirunyan:2018lul,Sirunyan:2020eab,CMS:2021cox,CMS:2021few}
Collaborations.
Most of these searches have been performed in events with at least one lepton.
The present search is performed in fully hadronic final states, which feature larger branching fractions but larger backgrounds.
Previous searches in fully hadronic final states have been performed for electroweak production of SUSY in the di-Higgs and other diboson
channels by the ATLAS~\cite{ATLAS:2018tti,ATLAS:2021yqv} and CMS~\cite{CMS:2014kxj,CMS:2017nin} Collaborations.
In this search we utilize machine learning algorithms to identify hadronically decaying high transverse momentum (\pt) \PW, \PZ, and Higgs bosons
reconstructed as large-radius jets~\cite{Sirunyan:2020lcu}.
The search uses a sample of LHC proton-proton ($\Pp\Pp$) collisions at $\sqrt{s} = 13\TeV$ collected by the CMS experiment
between 2016 and 2018, corresponding to an integrated luminosity of \fullLumi.
Tabulated results are provided in the HEPData record for this search~\cite{hepdata}.

\section{Simulated event samples}
\label{sec:samples}

Simulations of SM processes are used for optimizing selection criteria and computing several correction factors.
These factors are used to predict the rates of SM backgrounds based on observations in various control regions.
The production of \ttjets, \wjets, \zjets, Drell--Yan, and quantum chromodynamics (QCD) multijet events is simulated at leading order (LO) using the \MGvATNLO (2.2.2 for 2016 and 2.4.2 for 2017--2018) generator~\cite{Alwall:2014hca}.
The \ttjets events are simulated with up to three additional partons at the matrix-element level, while the other samples are simulated with up to four additional partons at the matrix-element level.
Single top quark events for all channels are modeled at next-to-leading order (NLO) in perturbative QCD.
The \MGvATNLO generator is used
for $s$-channel production of single top quark events, while \POWHEG v1.0 (v2.0)~\cite{Nason:2004rx,Frixione:2007vw,Alioli:2010xd,Alioli:2009je,Re:2010bp} is
used to simulate $t$-channel and associated $\PQt\PW$ production for 2016 (2017--2018).  Additional small backgrounds, such as \ttbar produced in association
with one or more SM bosons, are similarly produced at NLO with either \MGvATNLO or \POWHEG~\cite{Melia:2011tj,Nason:2013ydw,Hartanto:2015uka}.
Events for 2016 (2017--2018) are generated using the NNPDF~3.0~(3.1)~\cite{Ball:2014uwa,Ball:2017nwa} set of parton distribution functions (PDFs).
Parton showering and fragmentation are performed using the \PYTHIA~8.2~\cite{Sjostrand:2014zea} program with
the underlying event models detailed in Refs.~\cite{Khachatryan:2015pea,Sirunyan:2019dfx}.  The detector simulation is performed
with \GEANTfour~\cite{Agostinelli:2002hh}. Normalization of the simulated background samples is performed using the most accurate cross section calculations available~\cite{Alioli:2009je,Re:2010bp,Alwall:2014hca,Gavin:2012sy,Gavin:2010az,Li:2012wna,Gehrmann:2014fva,Campbell:2011bn,Beneke:2011mq, Cacciari:2011hy, Baernreuther:2012ws,Czakon:2012zr,Czakon:2012pz,Czakon:2013goa,Czakon:2011xx,Kulesza:2018tqz,deFlorian:2016spz}, which typically correspond to NLO or next-to-NLO accuracy.
Samples that are simulated at NLO with \MGvATNLO adopt the \textsc{FxFx} scheme~\cite{Frederix:2012ps} for matching partons from the matrix-element calculation to those from parton showers. Samples simulated at LO adopt the MLM scheme~\cite{Alwall:2007fs} for the same purpose.

Signal events are generated with the \MGvATNLO generator at LO precision
in a similar manner to the SM backgrounds, with up to two additional partons at the matrix-element level.
The detector simulation of signal samples is performed with the CMS fast simulation package~\cite{Abdullin:2011zz,Giammanco:2014bza}.
The signal samples are corrected for differences with respect to the \GEANTfour-based simulation.
Both the SM background samples and the signal samples are generated with nominal distributions
of additional $\Pp\Pp$ interactions per bunch crossing, referred to as pileup, which are then rescaled to match the observed pileup distribution.
The cross sections used for normalizing the signal yields are computed at NLO plus next-to-leading logarithmic precision~\cite{Beenakker:1999xh,Debove:2010kf,Fuks:2012qx,Fuks:2013vua,Fiaschi:2018hgm}.

\begin{figure*}[ht!]
\centering
\includegraphics[width=0.32\textwidth]{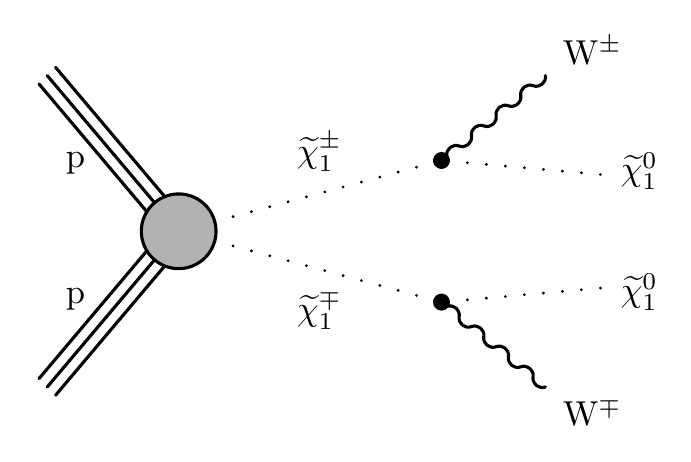}
\includegraphics[width=0.32\textwidth]{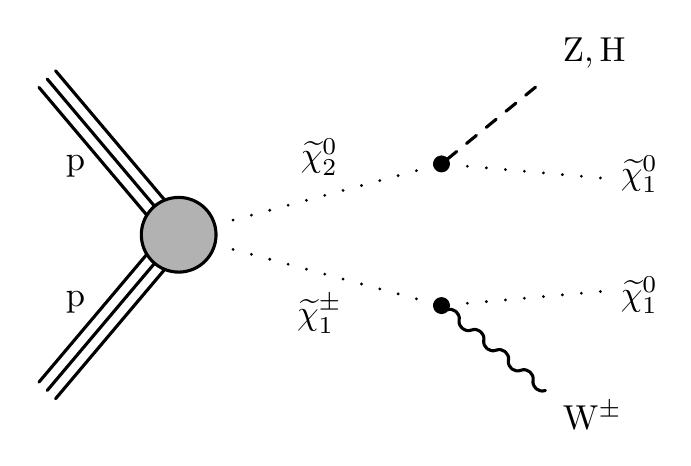}
\includegraphics[width=0.32\textwidth]{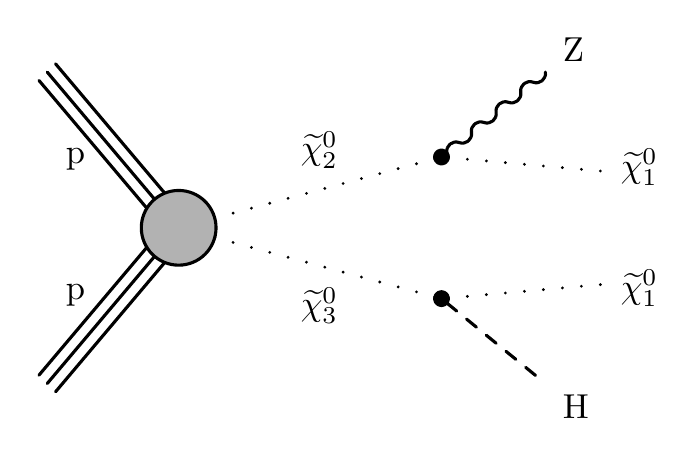}
\caption{Diagram for the production of \firstchargpair with the \firstcharg decaying to a \PW boson and \firstchi (left). In the \chargneut production diagram (middle) the \firstcharg decays to a \PW boson and the \secondchi decays to either a \PZ boson or a Higgs boson and \firstchi. In the case of \neuttwoneutthree prduction (right) the \secondchi and \thirdchi decay to a \PZ boson and Higgs boson, respectively, along with a \firstchi each.}
\label{fig:diagrams}
\end{figure*}

We interpret our results in terms of simplified models~\cite{ArkaniHamed:2007fw,Alwall:2008ag,Alwall:2008va,Alves:2011wf,Chatrchyan:2013sza}
with the assumption that a chargino pair \firstchargpair or a chargino-neutralino pair \chargneut
is produced.
In the simplified model scenarios, the chargino always decays to a \PW boson and the \firstchi, while the neutralinos decay 100\% of the time to either a \PZ or Higgs boson plus the \firstchi, where the \firstchi is assumed to be the lightest supersymmetric particle (LSP),
as depicted in Fig.~\ref{fig:diagrams}.
The simplified models are ``TChiWW'', \firstchargpair production with \firstcharg decaying to a \PW boson and \firstchi;
``TChiWZ'', \chargneut production with \secondchi decaying to a \PZ boson and \firstchi;
and ``TChiWH'', \chargneut production with \secondchi decaying to a Higgs boson and \firstchi.
In the TChiWZ and TChiWH scenarios, the \firstcharg and \secondchi are the mass-degenerate next-to-lightest supersymmetric particles (NLSPs),
while in the TChiWW scenario, the \firstcharg is the NLSP.
For these scenarios, we assume the \firstcharg and \secondchi to be purely wino-like.

In addition to these simplified models, we also consider two more specific scenarios in which all of the winos or all of the higgsinos are mass degenerate, which we refer to as the wino- and higgsino-like NLSP scenarios.
For the wino-like NLSP scenario, we assume a mass-degenerate \chargneut pair with a
lighter bino-like \firstchi as the LSP.
In this scenario, both \firstchargpair and \chargneut production are expected and would contribute to the targeted signatures, so these two production processes are combined and searched for simultaneously.
We consider two options for the decay of the \secondchi: either $\secondchi\to\PZ\firstchi$ or $\secondchi\to\PH\firstchi$, with 100\% branching fraction in each case.
For the higgsino-like NLSP scenario, we assume mass-degenerate \firstcharg, \secondchi, and \thirdchi with a lighter bino-like \firstchi.
In this scenario, the production of \firstchargpair, \chargneut, \chargneutthree, and \neuttwoneutthree are combined and searched for simultaneously.
We assume 100\% branching fractions of $\secondchi\to\PZ\firstchi$ and $\thirdchi\to\PH\firstchi$ as expected for
scenarios with $\tan\beta\sim 1$~\cite{Han:2013kza,Agrawal:2017zwh}.
We consider signal events with \WW, \WZ, \WH, or \ZH and large \ptmiss.
The \WW, \WZ, and \WH events arise from \firstchargpair, \chargneut, or \chargneutthree production.
The \ZH events arise from \neuttwoneutthree production as shown in Fig.~\ref{fig:diagrams} (right).
All the other sparticles are assumed to be too heavy to affect the production of charginos and neutralinos~\cite{Fuks:2012qx,Fuks:2013vua}.

\section{The CMS detector and event reconstruction}
\label{sec:reconstruction}

A detailed description of the CMS detector, together with a definition of the coordinate system used
and the relevant kinematic variables, can be found in Ref.~\cite{Chatrchyan:2008zzk}. Briefly,
a superconducting solenoid of 6\unit{m} internal diameter provides a magnetic field of 3.8\unit{T}. Within
the solenoid volume are a silicon pixel and strip tracker covering pseudorapidity $\abs{\eta} < 2.5$, a lead tungstate crystal electromagnetic
calorimeter, and a brass and scintillator hadron calorimeter, each composed of a barrel
and two endcap sections and covering $\abs{\eta} < 3$. Forward calorimeters extend the pseudorapidity coverage to include $3 < \abs{\eta} < 5.2$. Muons are detected in gas-ionization chambers embedded in the steel flux-return yoke
outside the solenoid.

The particle-flow (PF) algorithm~\cite{CMS-PRF-14-001}
combines all detector subsystems to identify and reconstruct charged and neutral hadrons,
photons, electrons, and muons.
This analysis utilizes PF jets, which are produced by clustering
PF candidates into anti-\kt~\cite{Cacciari:2008gp,Cacciari:2011ma} jets.
We utilize two jet collections
in which the clustering size is either 0.4 (AK4 jets) or 0.8
(AK8 jets).  The latter is well-suited for reconstructing hadronic decays of \PW, \PZ, and Higgs
bosons with $\pt > 200\GeV$.
We correct the jets for pileup effects~\cite{Sirunyan:2020foa,Bertolini:2014bba} and the jet energy scale~\cite{JEC}.
The AK8 jet mass \mJ is calculated using the soft-drop algorithm~\cite{softdrop}.
We consider only AK4 jets with $\pt>30\GeV$ and $\abs{\eta}<2.4$ and AK8 jets with
$\pt>200\GeV$ and $\abs{\eta}<2$ that satisfy a set of quality criteria
to eliminate jets from spurious sources, such as electronics noise~\cite{Sirunyan:2020foa}.

The variable \ptvecmiss is defined as the negative vector sum of \ptvec for
all PF candidates, and its magnitude is denoted by \ptmiss. Known detector effects are accounted for by adjusting
\ptmiss for jet energy corrections~\cite{Sirunyan_2019}.  A set of
quality criteria is used to identify and eliminate events in which detector noise, inoperable calorimeter
cells, beam halo, and other effects mimic \ptmiss~\cite{Sirunyan_2019}.

The identification of jets originating from a bottom quark (\PQb tagging) is performed by applying a version of the combined secondary vertex algorithm
that utilizes deep-learning techniques~\cite{Sirunyan:2017ezt}.  For the medium working point
utilized in this search, the efficiency of tagging \PQb jets with $\pt>30\GeV$, as measured in \ttbar events,
is about 68\%; the probability of misidentifying jets arising from the hadronization of a charm quark is roughly 12\% and for jets from a light-flavor
quark or a gluon it is roughly 1\%.

The lepton content of events is used to separate the fully hadronic signal samples from the single-lepton control samples, as described in Sections~\ref{sec:selections} and \ref{sec:bkg_est}.
We require electron and muon candidates~\cite{CMS:2015xaf,CMS:2018rym} to have $\pt>10\GeV$ and to lie within $\abs{\eta} < 2.5$ for electrons and $\abs{\eta} < 2.4$ for muons.
Electron candidates are required to satisfy the ``veto'' requirements described in Ref.~\cite{CMS:2015xaf}, and muon candidates are required to satisfy the ``medium'' requirements described in Ref.~\cite{CMS:2018rym}.
We also impose isolation
requirements on electron and muon candidates to suppress those arising from jets erroneously
identified as leptons, as well as genuine leptons from hadron decays, using the same criteria as in Refs.~\cite{CMS:2016krz,Sirunyan:2019ctn}.
To recover electrons or muons that fail the identification requirements, as well as \PGt leptons via their
one-prong hadronic decays, we make use of isolated tracks.
Isolated tracks are required to satisfy $\abs{\eta}<2.4$ and $\pt>5\GeV$ for PF electrons and muons,
and $\pt>10\GeV$ for PF hadrons~\cite{Sirunyan:2019ctn}.

Photon candidates, which are only used to remove events from the data set as described in Section~\ref{sec:selections}, are required to have ${\pt>100\GeV}$ and ${\abs{\eta}<2.4}$ and to be isolated from charged
hadrons, neutral hadrons, and electromagnetic particles~\cite{Khachatryan:2015iwa}.

Candidate AK8 jets consistent with the decay and subsequent fragmentation and hadronization of
heavy SM bosons are identified using a deep neural network (DNN).
The DNN is designed to classify AK8 jets arising from hadronically decaying particles of five main categories: \PW, \PZ, \PH, \PQt, and ``other''.
Jets are further classified into subcategories according to their
likelihood of originating from
specific decays of these unstable particles, \eg, $\PZ(\bbbar)$, $\PZ(\ccbar)$, etc.
The architecture of the DNN and details of its training can be found in Ref.~\cite{Sirunyan:2020lcu}.
Each classification has a corresponding score that is used to develop jet taggers.

In this analysis, three jet taggers, which were developed from these classifiers, are used to categorize AK8 jets: the \PW tagger, the \PV tagger, and the \bbbar tagger.
These taggers are non-exclusive, \ie, a single AK8 jet may be tagged by any, all, or none of the three taggers.
Each tagger involves the DNN and the jet mass.
The \PW tagger identifies jets consistent with a \Wqqprime decay by their mass
($65<\mJ<105\GeV$) and a DNN score optimized for distinguishing between
hadronic \PW boson decays and QCD jets.
The \PV tagger identifies jets consistent with a \Wqqprime or $\PZ(\PQq\PAQq)$ decay by their mass
($65<\mJ<105\GeV$) and a DNN score optimized in a similar manner to the \PW tagger,
but utilizing adversarial training~\cite{Louppe:2016ylz} to decorrelate the DNN score
and the jet mass, which allows the \PV tagger to be sensitive to the hadronic decays of both the \PW and \PZ bosons despite a higher misidentification rate.
The \bbbar tagger identifies AK8 jets consistent with a $\PZ(\bbbar)$ or $\PH(\bbbar)$ decay by their mass
($75<\mJ<140\GeV$) and a DNN score optimized for identifying Lorentz-boosted \bbbar topologies.
The \bbbar tagger also utilizes adversarial training to decorrelate the DNN score and the jet mass.
The \bbbar tagger has an efficiency of about 54\% for jets with $\pt>300\GeV$ originating from Higgs or \PZ bosons decaying into \bbbar, and
a misidentification rate of roughly 2.5\% for QCD jets.
Similarly, the \PW tagger has an efficiency of about 41\% for jets arising from hadronic \PW boson decays and a misidentification rate of roughly 1\%,
and the \PV tagger has an efficiency of about 45\% for tagging hadronic \PW and \PZ boson decays and a misidentification rate of roughly 2.5\%.

\section{Triggers and event selection}
\label{sec:selections}
Candidate events are recorded based on a trigger~\cite{Khachatryan:2016bia} that requires \ptmiss to be larger than a time-dependent threshold that varies between 90 and 140\GeV.
The efficiency of these triggers is measured using
a separate data set in which events are recorded based on the requirement that a single electron or
single muon is reconstructed.
For \ptmiss of about 200\GeV, which is the minimum \ptmiss used in this search,
the trigger efficiency is found to be 95, 78, and 74\% for the 2016, 2017, and 2018 data-taking periods, respectively.
This variation in the trigger efficiency is caused by the changes in the trigger threshold over time, but the efficiency rises above 98\% for $\ptmiss>270\GeV$ for all years.
When computing event yields, parameterizations of these efficiencies
are used to correct simulations.
For ancillary measurements of jet tagging rates in the data, a combination of single-lepton and dilepton triggers is used.

All events in our signal regions (SRs) are
required to satisfy a common set of baseline selection criteria.  Each event
is required to have a primary vertex~\cite{CMS:2014pgm} and no isolated leptons, isolated photons, or isolated tracks.
Leptons, in particular, typically arise from a \PW boson decay, which constitutes a
major background.
The requirement of no leptons in our SRs makes this analysis orthogonal to other searches targeting common signals
in final states that include leptons~\cite{Sirunyan:2017zss,Sirunyan:2017lae,Sirunyan:2018ubx,Sirunyan:2018lul,Sirunyan:2020eab}.
Events containing isolated tracks are only removed from the data set if the transverse mass based on the track satisfies $\mT = \sqrt{\smash[b]{2\pt \ptmiss (1-\cos\Delta\phi)}}<100\GeV$,
where $\Delta\phi$ refers to the azimuthal separation between the track and \ptvecmiss directions.
Low \mT is typical of tracks from $\PW \to \Pell\PGn$ decays but less common in signal
events with isolated tracks.
Events are required to have $\ptmiss>200\GeV$ and $\HT>300\GeV$, where \HT is defined
as the scalar sum of \pt for all AK4 jets which satisfy the kinematic and quality criteria mentioned in Section~\ref{sec:reconstruction}.
Large \ptmiss and \HT values are typical of chargino and neutralino production when a high-momentum boson is present.
Signal events typically have two AK8 jets and four AK4 jets (note that AK4 jets may overlap AK8 jets); accordingly we require at least two AK8 jets and 2--6 (inclusive) AK4 jets.
We also impose requirements on the azimuthal angle $\Delta\phi_\mathrm{i}$ between \ptvecmiss and each of the four highest \pt AK4 jets,
where the subscript $\mathrm{i}$ refers to the \pt ordering of the jets.
Each event must satisfy $\Delta\phi_{1,2,3,4}>1.5,0.5,0.3$, and 0.3.
Events must satisfy $\Delta\Phi_{1,2}>1.5$ and 0.5, where $\Delta\Phi_\mathrm{i}$ is defined analogously to $\Delta\phi_\mathrm{i}$ using the two highest \pt AK8 jets.
These requirements on azimuthal angles suppress the background from QCD multijet events, for which \ptvecmiss is usually aligned along a jet direction.
Within this baseline phase space, four SRs are defined.
One SR, which we refer to as the \PQb{}-veto region, requires zero \PQb{}-tagged AK4 jets ($\nb=0$) and targets signal events in which SM bosons decay to light-flavor quarks.
The remaining three SRs, which we refer to as the \PQb{}-tag regions, require at least one \PQb{}-tagged AK4 jet ($\nb\geq1$) and target signal events in which at least one SM boson decays to \bbbar.
In addition to these SRs, we also define several control regions (CRs), which are used to help constrain the background estimates, as described in Section~\ref{sec:bkg_est}.

\subsection{The \texorpdfstring{\PQb{}}{b}-veto search region}
\label{sec:bvetoSR}

The \PQb{}-veto SR seeks to isolate events that are consistent with the production of \WW or \WZ pairs of bosons, along with a large \ptmiss.
In addition to the baseline event selection described above, the \PQb{}-veto SR requires that at least two AK8 jets satisfy $65 < \mJ < 105\GeV$.
At least one AK8 jet must be \PW tagged, and at least one other AK8 jet must be \PV tagged, as described in Section~\ref{sec:reconstruction}.
A summary of the \PQb{}-veto SR requirements is shown in Table~\ref{tab:bveto_regions}.
The \PQb{}-veto SR is further subdivided into nine bins of \ptmiss.
The lower \ptmiss bin boundaries are 200, 250, 300, 350, 400, 450, 500, 600, and 800\GeV.

\begin{table}[ht!]
	\centering
	\topcaption{Summary of tagging requirements for the \PQb{}-veto SR and CRs.
		Each of these regions includes the baseline selection described in Section~\ref{sec:selections}
		and requires zero \PQb{}-tagged AK4 jets and at least two AK8 jets satisfying $65 < \mJ < 105\GeV$.
		The SR and CRs are described in detail in Sections~\ref{sec:bvetoSR} and \ref{sec:bkg_est_bveto}, respectively.
		The \PW and \PV taggers are described in Section~\ref{sec:reconstruction}.}
	\label{tab:bveto_regions}
	\begin{tabular}{cc}
		Region                                & Requirements                          \\
		\hline
		\multirow{3}{*}{\PQb{}-veto SR}       & ${\geq}1$ \PV{}-tagged jet            \\
		                                      & ${\geq}1$ \PW{}-tagged jet            \\
		                                      & ${\geq}2$ \PV{}- or \PW{}-tagged jets \\[\cmsTabSkip]
		\multirow{2}{*}{\PQb{}-veto 0-tag CR} & 0 \PV{}-tagged jets                   \\
		                                      & 0 \PW{}-tagged jets                   \\[\cmsTabSkip]
		\multirow{2}{*}{\PQb{}-veto 1-tag CR} & 1 \PV{}-tagged jet                    \\
		                                      & 0 other \PW{}-tagged jets             \\[\cmsTabSkip]
	\end{tabular}
\end{table}

The main background in the \PQb{}-veto SR arises from \vjets production with $\PW\to\Pell\PGn$ or $\PZ\to\PGn\PAGn$.
The $\PW\to\Pell\PGn$ background is substantially reduced by requiring the number of reconstructed charged leptons and isolated tracks to be zero.
These events still satisfy the event selection criteria when the
charged lepton lies outside the lepton acceptance, is not reconstructed, or is not isolated.
In \vjets background events, both \PW{}- and \PV{}-tagged AK8 jets arise from misidentification of jets not originating from hadronic \PW or \PZ boson decays.
These events together with background events arising from QCD multijet production do not contain any resonance reconstructed as a single AK8 jet, and are referred to as the ``\zerores'' background.

The next largest background contributions come from \ttbar, single top quark, and diboson production.
These events typically have one leptonically decaying vector boson and one hadronically decaying vector boson.
Therefore, one \PW{}- or \PV{}-tagged AK8 jet arises from a hadronic \PW or \PZ boson decay, and the other tag arises from misidentification.
We refer to these backgrounds, which contain only one resonance reconstructed as a single AK8 jet, as the ``\oneres'' backgrounds.
The remaining minor background contributions, which constitute less than 10\% of the expected event yield in any \ptmiss bin of the SR, are expected from rare processes such as triboson production
and \ttbar pairs produced in association with a \PW, \PZ, or Higgs boson.

\begin{figure}[ht!]
	\centering
	\includegraphics[width=0.49\textwidth]{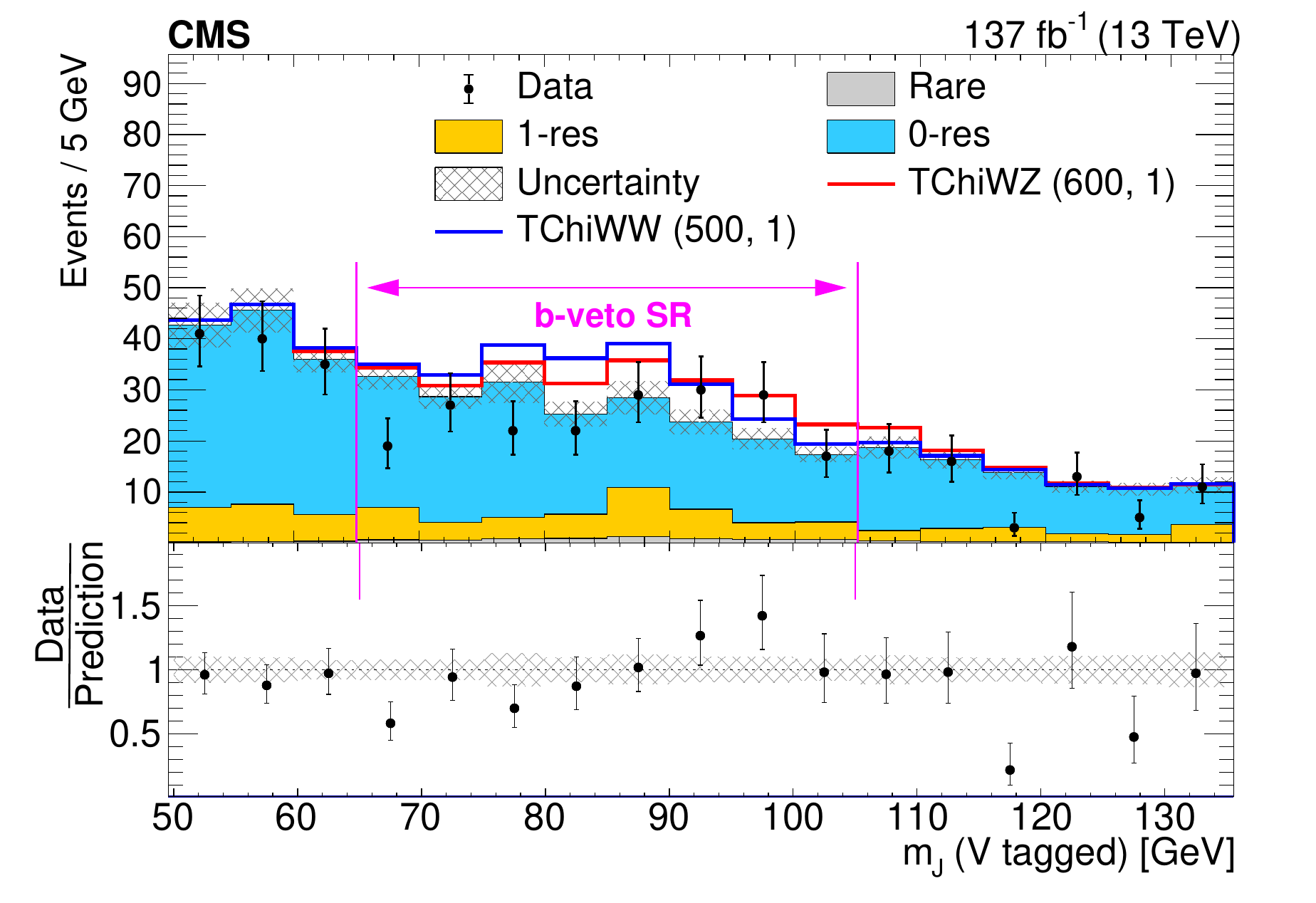}
	\includegraphics[width=0.49\textwidth]{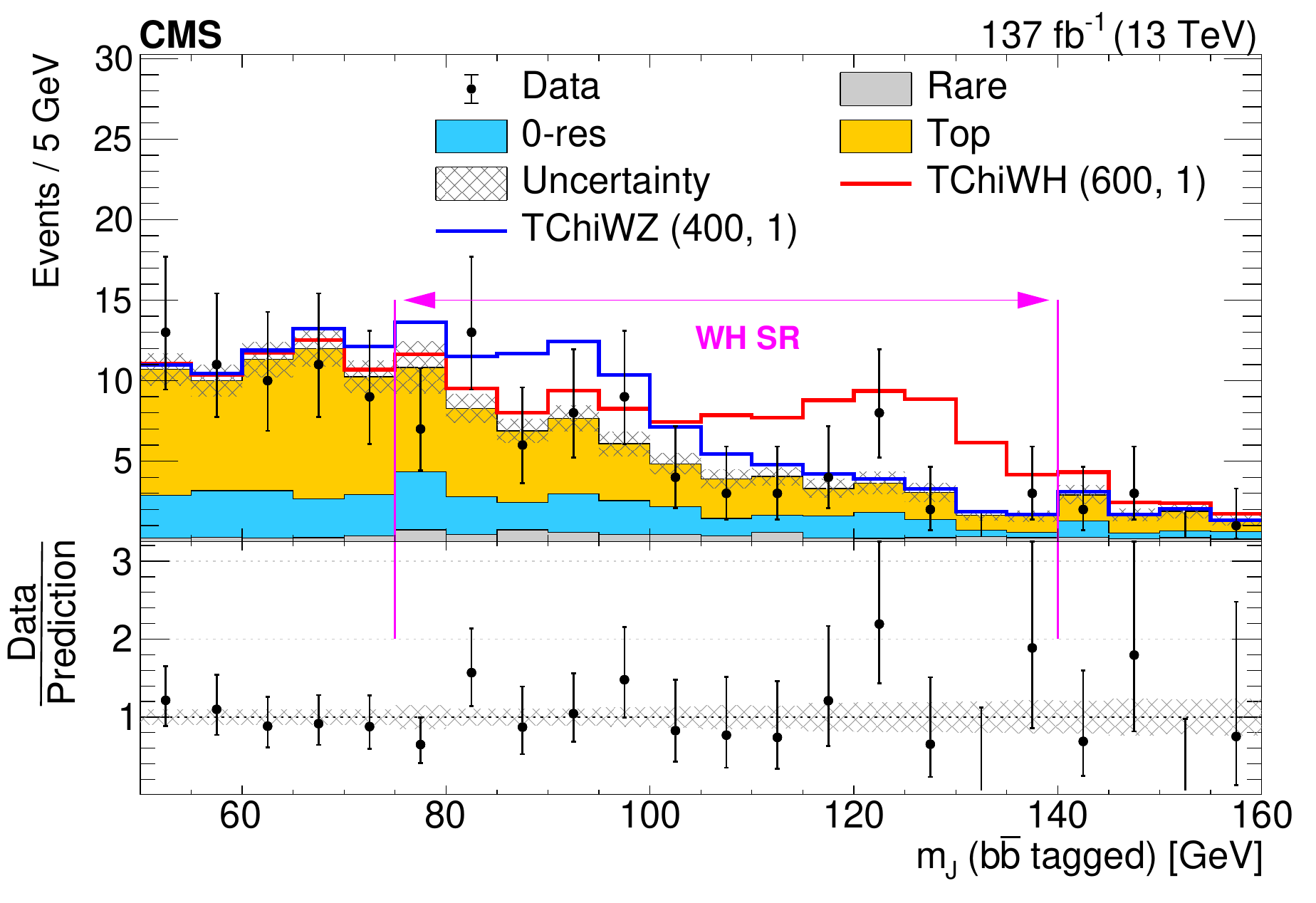}
	\caption{Distributions of the jet mass for \PV{}-tagged AK8 jets in the \PQb{}-veto SR (\cmsLeft) and \bbbar-tagged AK8 jets
		in the \WH SR (\cmsRight).
		The jet mass requirements for the \PV and \bbbar taggers have been loosened in these figures.
		The filled histograms show the SM background simulation, scaled such that the yield within the SR matches the total SM background predictions.
		The open histograms show the sum of the scaled SM background simulations and of the expectations for selected signal models, which are
		denoted in the legend by the name of the model followed by the assumed masses
		of the NLSP and LSP in \GeVns. The observed event yields are indicated by black markers.
		The hatched gray bands correspond to the statistical uncertainties in the SM predictions, but no systematic uncertainties are included.}
\label{fig:mj_dist}
\end{figure}

The \mJ distribution of \PV{}-tagged jets for signal, expected SM backgrounds, and observed event yields
in the \PQb{}-veto SR is shown in Fig.~\ref{fig:mj_dist}~(\cmsLeft), in which the mass requirement of the \PV tagger has been loosened to show the behavior of the signal and backgrounds in both the tagged regions and the sidebands.
In some events, there are multiple \PV{}-tagged jets, some of which could also be \PW tagged.
In such events, the jet that is both \PW and \PV tagged and has the highest \PW tagger DNN score is ignored,
and the \PV{}-tagged jet with the highest \PV tagger DNN score among the remaining jets is plotted.
The simulated event yields are scaled such that the yield within the SR matches our total SM background predictions, which will
be described in Section~\ref{sec:bkg_est}.
The distributions for the TChiWZ and TChiWW signals show peaks in the region of 65--105\GeV, corresponding to the mass requirement of the \PV tagger.

\subsection{The \texorpdfstring{\PQb{}}{b}-tag search regions}
\label{sec:btagSR}

To maximize our acceptance of signal events, we define three SRs for events containing at least one \PQb{}-tagged AK4 jet.
These SRs are defined in terms of the numbers of ``\PW boson candidates'' and ``Higgs boson candidates'', which are AK8 jets selected according to \mJ and according to their proximity to a \PQb{}-tagged AK4 jet.
An AK8 jet is said to contain a \PQb{}-tagged AK4 jet if the event includes at least one \PQb{}-tagged AK4 jet that satisfies $\Delta R(\PQb{}\text{ jet}, \text{AK8 jet}) < 0.8$, where $\Delta R = \sqrt{\smash[b]{\left(\Delta \eta\right)^2 + \left(\Delta \phi\right)^2}}$ is the distance in the pseudorapidity-azimuth plane.
A \PW boson candidate is any AK8 jet that lies in the \PW{}/\PZ mass window $65 < \mJ < 105\GeV$ and does not contain a \PQb{}-tagged jet.
Similarly, a Higgs boson candidate is any AK8 jet that lies in the \PZ{}/\PH mass window $75 < \mJ < 140\GeV$ and contains at least one \PQb{}-tagged jet.
These two classes of AK8 jets do not overlap, but not every AK8 jet falls into either class.
We further subdivide these classes based on jet tagging.
The \PW tagger is used to divide \PW boson candidates into tagged and untagged subcategories, and the \bbbar tagger is similarly used to divide Higgs boson candidates into tagged and untagged subcategories.

The three SRs with \PQb{}-tagged jets are referred to as the \WH, \PW, and \PH SRs.
These SRs isolate events in which a \WH, \WZ, or \ZH pair of bosons is produced with large \ptmiss.
The \WH SR requires at least one tagged \PW boson candidate and at least one tagged Higgs boson candidate.
The \PW SR requires at least one tagged \PW boson candidate and exactly zero tagged Higgs boson candidates, although untagged Higgs boson candidates are permitted.
The \PH SR requires at least one tagged Higgs boson candidate and exactly zero tagged \PW boson candidates, but similarly imposes no requirement on untagged \PW boson candidates.
These requirements are summarized in Table~\ref{tab:btag_regions}.

The SM backgrounds that
contribute to the three \PQb{}-tag SRs are top quark pair production, especially at low \ptmiss,
and \vjets events, especially at high \ptmiss.  In all three \PQb{}-tag SRs, there
are small contributions from diboson and triboson production, as well as \ttbar production in association
with a \PW, \PZ, or Higgs boson.
The backgrounds containing top quarks include a mixture of events with one and two resonances each reconstructed as a single AK8 jet, and so we refer to these backgrounds in the \PQb{}-tag region as ``top'' rather than as ``\oneres''.

\begin{table*}
	\centering
	\topcaption{Summary of tagging requirements for the \PQb{}-tag SRs and CRs.
		Each of these regions includes the baseline requirements described in Section~\ref{sec:selections} and requires at least one \PQb{}-tagged AK4 jet and at least two AK8 jets.
		The SRs and CRs are described in detail in Sections~\ref{sec:btagSR} and \ref{sec:bkg_est_btag}, respectively.
		The \bbbar and \PW taggers are described in Section~\ref{sec:reconstruction}, and the definitions of
		\PW and Higgs boson candidates are given in Section~\ref{sec:btagSR}.
		In addition to the six regions described in this table, the \PQb{}-tag predictions also use six single-lepton CRs
		that are identical except that exactly one charged lepton is required.
		A dash (\NA) indicates that no requirement is imposed.}
	\label{tab:btag_regions}
	\begin{tabular}{ccccc}
		               & \multicolumn{2}{c}{\PW boson candidate} & \multicolumn{2}{c}{Higgs boson candidate} \\
		               & \PW tagged & not \PW tagged             & \bbbar tagged & not \bbbar tagged         \\
		\hline
		\WH SR         & ${\geq}1$  & \NA                        & ${\geq}1$     & \NA                       \\
		\PW SR         & ${\geq}1$  & \NA                        & 0             & \NA                       \\
		\PH SR         & 0          & \NA                        & ${\geq}1$     & \NA                       \\
		\WH antitag CR & 0          & ${\geq}1$                  & 0             & ${\geq}1$                 \\
		\PW antitag CR & 0          & ${\geq}1$                  & 0             & 0                         \\
		\PH antitag CR & 0          & 0                          & 0             & ${\geq}1$                 \\
	\end{tabular}
\end{table*}

The \PQb{}-tag SRs are also subdivided into bins of \ptmiss.
The \PW and \PH SRs use the same \ptmiss binning as the \PQb{}-veto SR.
For the \WH SR, the last two \ptmiss bins are merged because of the small expected number
of events in the corresponding CRs.

The distribution of the Higgs boson candidate \mJ for events in the \WH SR
for signal, expected SM backgrounds, and observed event yields is shown in Fig.~\ref{fig:mj_dist}~(\cmsRight), in which the
Higgs boson candidate mass requirement has been loosened by $\sim$20\GeV on either side to show the distribution both in and near the SR.
The predicted yields
from simulation are scaled such that the total SM background yields in the SR match the total prediction using the procedure described in Section~\ref{sec:bkg_est}.
Distributions for the TChiWZ and TChiWH signals show peaks in the region of 75--140\GeV, corresponding to the \PZ and Higgs boson masses.

\section{Background estimation}
\label{sec:bkg_est}

To constrain the contribution from the dominant SM processes in each of our SRs, several CRs
are defined.  Each CR has reduced contributions from signal processes and enhanced contributions from
SM background processes.
Most of our CRs isolate different background components by inverting
various tagging requirements to select events with a misidentified boson.  For the \PQb{}-tag SRs, where
backgrounds involving top quarks are more prominent, a combination of CRs with inverted tagging requirements and with one charged lepton is used.

\subsection{Background estimation for the \texorpdfstring{\PQb}{b}-veto search region}
\label{sec:bkg_est_bveto}

The sum of 0- and \oneres background events in the \PQb{}-veto SR is estimated simultaneously from two mutually-exclusive CRs.
Both CRs require at least two AK8 jets that satisfy the \WZ mass requirements.
The 0-tag CR is defined by requiring that all AK8 jets satisfying the \WZ mass requirements are neither \PW tagged nor \PV tagged.
This CR is dominated by the \zerores background.
The 1-tag CR requires exactly one \PV{}-tagged jet, and requires that all other AK8 jets that satisfy the \WZ mass requirements are neither \PW tagged nor \PV tagged.
The main contribution to the 1-tag CR also comes from the \zerores background, but this CR has increased contributions from the \oneres background compared to the 0-tag CR.
The requirements for both \PQb{}-veto CRs are summarized in Table~\ref{tab:bveto_regions}.

The background yields in the \PQb{}-veto SR are estimated using two sets of transfer factors derived from simulation, $\mathcal{R}_\mathrm{i}$,
defined as the ratio of the summed 0- and \oneres event yields in the
SR divided by those in either the 0-tag ($\mathcal{R}_0$) or 1-tag CR ($\mathcal{R}_1$).
The values of $\mathcal{R}_\mathrm{i}$ are computed separately for each \ptmiss
bin and typically range between 0.2 and 0.3.
The contributions of rare processes to the SR and CRs are taken from simulation with appropriate data-to-simulation corrections applied~\cite{Sirunyan:2020lcu,Sirunyan:2017ezt,JEC}.
The total background prediction is given by
\begin{linenomath}
\begin{equation*}
N^\text{pred}_\text{SR} = \mathcal{R}_\mathrm{i} ( N^\text{data}_{\text{CR}_\mathrm{i}} - N^\text{MC}_{\text{CR}_\mathrm{i},\text{rare}} )
+ N^\text{MC}_\text{SR,rare},
\end{equation*}
\end{linenomath}
where $N$ denotes the number of events expected (or observed in data) in regions and from processes indicated by the superscripts and subscripts, $\mathcal{R}_\mathrm{i} = N^\text{MC}_\text{SR,0\&\oneres} / N^\text{MC}_{\text{CR}_\mathrm{i},\text{0\&\oneres}}$,
and $\text{CR}_\mathrm{i}$ is either the 0- or 1-tag CR. The final background predictions for the SR are determined by a simultaneous fit of the two CRs.
The background composition of the 1-tag CR is very similar to the \PQb{}-veto SR, while the signal contamination is smaller in the 0-tag CR.
The level of signal contamination in the CRs depends on the choice of SUSY model and on the assumed NLSP and LSP masses.
Both CRs are included in the fits to the data described in Section~\ref{sec:results}, and the signal contamination, although small, is taken into account in the fits.
Using both CRs allows us to benefit from the varying background compositions and signal contamination levels.

The \PW and \PV tagging rates in simulation are corrected to match those measured in data.
The corrections for jets matched to generator-level \PW bosons are obtained from a \ttbar{}-enriched sample in which one of the \PW bosons from the top quark decays leptonically and the other decays hadronically~\cite{Sirunyan:2020lcu}.
The same corrections are also applied to jets matched to generator-level \PZ bosons.
The corrections for misidentified jets are obtained from a sample of events with an $\HepParticle{\Pell}{}{+}\HepParticle{\Pell}{}{-}$ ($\Pell=\Pe$, \PGm) pair and at least one AK8 jet.
The \PW{}- or \PV{}-tagged jets in these events are dominated by misidentification.
Uncertainties associated with the determination of
these corrections are propagated to the final predictions.

Our \PQb{}-veto background prediction method relies on the modeling of the AK8 jet \PW{}- and \PV{}-tagging efficiencies in simulation after the corrections described above are applied.
Several validation regions (VRs) are defined to check how well the simulation models the tagging efficiencies in the phase space close to our SR and CRs.
For one set of VRs, events are required to satisfy the baseline event selection criteria except for the
second AK8 jet requirement,
and to have only one AK8 jet, which is \PW or \PV tagged.
These VRs are referred to as the 1-jet \PW{}- and \PV{}-tag VRs, respectively.
The predicted SM backgrounds, with the exception of the rare backgrounds, in each VR are based
on an extrapolation from CRs defined by the inversion of the DNN requirement for the \PW or \PV tag compared to the corresponding VR.
The predicted yields of the rare backgrounds are obtained directly from simulation, not via extrapolation from CRs.
The predictions in these 1-jet VRs are found to be compatible with our observations.

Similar tests are also performed using events that satisfy the baseline event selection criteria including the requirement of having at least two AK8 jets
but only one of them satisfying $65<\mJ<105\GeV$.
These events are orthogonal to events in any of \PQb{}-veto SR and CRs, as the \PQb{}-veto SR and CRs require at least two jets to satisfy $65<\mJ<105\GeV$.
We then require that only one AK8 jet is \PW or \PV tagged.
These \PQb{}-veto VRs are referred to as the 2-jet \PW{}- and \PV{}-tag VRs, respectively.
Predictions of the background yields, with the exception of the rare backgrounds, are made by extrapolating from CRs defined by inversion of the tagging DNN requirement.
As before the rare backgrounds yields are taken directly from simulation.

In the 2-jet \PW{}- and \PV{}-tag VRs,
the observed event yields are higher than the SM background predictions, primarily in the intermediate- to high-\ptmiss region,
which may be attributed to a dependence of the misidentification rate corrections on event topology or on AK8 jet multiplicity in this phase space.
We correct the background predictions in the \PQb{}-veto SR to account for these discrepancies.
The correction from the \PW{}-tag VR is applied to the extrapolation from the 1-tag CR to the SR,
and the corrections from the \PW{}- and \PV{}-tag CRs are both applied to the extrapolation from the 0-tag CR to the SR.
The corrections obtained from the 2-jet \PW{}- and \PV{}-tag VRs range from 1.03 and 1.01 at low \ptmiss to 1.48 and 1.27 in the highest \ptmiss bin, respectively.
The full size of these corrections is considered as a systematic uncertainty.

\subsection{Background estimation for the \texorpdfstring{\PQb}{b}-tag search regions}
\label{sec:bkg_est_btag}

For each of the three \PQb{}-tag SRs, we define three sets of CRs, for a total of nine CRs.
The first set, the antitag CRs, inverts the boson tagging requirements of the corresponding SRs.
In the antitag CRs, every Higgs boson candidate must not be \bbbar tagged, and every \PW boson candidate must not be \PW tagged.
Otherwise, the definitions of the \WH, \PW, and \PH antitag CRs, which are summarized in Table~\ref{tab:btag_regions}, are identical to the corresponding SRs.
The antitag CRs are enhanced in \vjets but still have contributions from events containing top quarks.

The two remaining sets of CRs are defined identically to the SRs and antitag CRs but require exactly one lepton, \ie, there is a
set of three single-lepton (\oneell) tagged CRs and a set of three \oneell antitag CRs. The \oneell CRs
feature enhanced rates of top quark pair production.  A summary of the selection for each SR and corresponding CR is
given in Table~\ref{tab:btag_regions}.
All of the SRs and CRs are mutually non-overlapping.

The \oneell CRs are used to constrain the estimates of top quark backgrounds in the SRs and zero-lepton antitag CRs, while the antitag CRs are used to constrain the estimates of the \zerores backgrounds in the SRs and tagged CRs.
The signal contamination in the \PQb{}-tag CRs is negligible in most of the CRs for most signal models.

The \oneell CRs are dominated by top quark pair production, with small contributions
from \wjets and even smaller contributions from single top quark production and
\ttH.
A transfer factor, $\mathcal{R}_{\zeroell/\oneell}$, is used to provide an
estimate of the number of top quark background events in either the SR or the \zeroell antitag CR, where \zeroell refers to the SRs and the antitag CRs with zero leptons.
The values of $\mathcal{R}_{\zeroell/\oneell}$ are computed from simulation, including all
corrections to the lepton reconstruction efficiencies, \PQb tagging efficiencies, and AK8 jet tagging efficiencies.

The predicted number of top quark background events in either the SR or the \zeroell antitag CR is given by
\begin{linenomath}
\begin{equation}
\label{eq:reson_pred}
N^{\text{pred},\zeroell}_{\mathrm{i},\text{top}} = \frac{N^{\text{MC},\zeroell}_{\mathrm{i},\text{top}}}{N^{\text{MC},\oneell}_{\mathrm{i},\text{all}}} N^{\text{data},\oneell}_\mathrm{i} = \mathcal{R}_{\zeroell/\oneell}N^{\text{data},\oneell}_\mathrm{i},
\end{equation}
\end{linenomath}
where $N^\text{MC}$ denotes the number of events expected from simulation, $N^\text{data}$ denotes the number of observed events, and $N^\text{pred}$ denotes the number of events predicted via this method.
Additionally, the subscript $\mathrm{i}$ denotes the tagging region, tag or antitag.
The subscript ``all'' refers to all of the SM backgrounds, while ``top'' refers to only the top quark backgrounds.

A transfer factor $\mathcal{R}_{\text{tag}}$ is used to constrain the \zerores event yields in the SR based on
observations in the \zeroell antitag CR.  The value of $\mathcal{R}_{\text{tag}}$ is computed using simulation and is the ratio of the number of \zerores events in the SR to the number of \zerores events in the
\zeroell antitag CR.  This transfer factor does not include the top quark backgrounds, which are constrained
separately via the \oneell CRs.  The transfer factor is corrected for differences in tagging rates between
simulation and data.  Corrections are derived separately for \PW{}- and \bbbar-tagged jets, and separately for
misidentified and correctly tagged jets~\cite{Sirunyan:2020lcu}.
Corrections for misidentified jets are derived in a similar manner to the corrections used for the \PQb{}-veto
regions, using samples of Drell--Yan events and requiring $\nb=1$.
All uncertainties in the tagging and misidentification rate corrections are propagated to the final predictions.

Using $\mathcal{R}_{\text{tag}}$ and the prediction of the top quark backgrounds described above,
the predicted \zerores background contribution to the SR is given by
\begin{linenomath}
\begin{equation}
\label{eq:nonreson_pred}
N^{\text{pred},\zeroell}_{\text{\zerores}} = \mathcal{R}_\text{tag}\left(N^{\text{data},\zeroell}_{\text{antitag}}-N^{\text{pred},\zeroell}_{\text{antitag,top}}-N^{\text{MC},\zeroell}_{\text{antitag,rare}}\right),
\end{equation}
\end{linenomath}
where $N^{\text{data},\zeroell}_{\text{antitag}}$ denotes the number of observed events in the \zeroell antitag CR,
$N^{\text{pred},\zeroell}_{\text{antitag,top}}$ denotes the predicted number of top quark background events from Eq.~\eqref{eq:reson_pred}, and
$N^{\text{MC},\zeroell}_{\text{antitag,rare}}$ denotes the number of rare background events, such as diboson and triboson events, expected from simulation.

The predictions of the top quark backgrounds and \zerores backgrounds are taken from Eqs.~\eqref{eq:reson_pred} and~\eqref{eq:nonreson_pred}, while the prediction of the rare backgrounds is taken from simulation.
These predictions are produced in each \ptmiss bin and in each
SR independently. Each transfer factor is derived separately for each \ptmiss bin except for
$\mathcal{R}_{\zeroell/\oneell}$, whose values above $\ptmiss>450$~(400)\GeV are averaged in the \PW and \PH (\WH) SRs.

The predictions of the SM backgrounds are tested in data using an orthogonal VR in which exactly
one AK8 jet is required. This VR is used to test the simulation-derived transfer factors,
misidentification scale factors, and simulation-based contamination terms in the CRs.
Within the VR, two pseudo-SRs are defined in which the AK8 jet is \PW tagged or \bbbar tagged, respectively.
For each pseudo-SR, a \zeroell antitag CR and \oneell tag and antitag CRs are defined for the background estimation.
All transfer factors are rederived in the VR phase space. Predictions
and observations in the pseudo-SRs are found to be statistically compatible.

\section{Systematic uncertainties}
\label{sec:syst}

For all SRs and CRs, the expected rate of signal and backgrounds is adjusted for
known differences between data and simulation in the jet tagging rates (using the \PQb, \bbbar, \PW,
and \PV taggers), lepton identification rates, and trigger efficiencies.
Additional corrections are applied to the predicted signal yields to account for known differences
between the \GEANTfour-based CMS detector simulation and the CMS fast simulation in
the AK8 jet mass, \PW tagging, \PV tagging, \bbbar tagging, \PQb tagging, and \ptmiss.
Uncertainties associated with the determination of these correction
factors are propagated to the final yield estimates, as is the uncertainty in the integrated luminosity~\cite{CMS-LUM-17-003,CMS-PAS-LUM-17-004,CMS-PAS-LUM-18-002}.
Uncertainties related to the determination of the trigger efficiency are
dominated by the limited size of the data sets in which we measure the trigger efficiency at high \ptmiss and by the differences between
the efficiencies determined using two classes of events, namely events with one electron and events with one muon, at low \ptmiss.
Uncertainties related to the corrections to the rates of jet misidentification are due to the limited
size of our dilepton data sets.
Uncertainties associated
with the lepton identification efficiency corrections and jet tagging rate corrections are detailed in Refs.~\cite{CMS:2018rym,CMS:2015xaf} and \cite{Sirunyan:2017ezt,Sirunyan:2020lcu},
respectively.
In this analysis, the \PW tagger is used to identify \Wqqprime decays,
but some signal events with \ZH and large \ptmiss in the higgsino-like NLSP scenario enter the \PQb{}-tag \WH or \PW SRs when $\PZ(\PQq\PAQq)$ decays are selected by the \PW tagger.
The \PW tagger efficiency is lower for $\PZ(\PQq\PAQq)$ decays than for \Wqqprime decays by 20--40\% in simulation,
but its performance is calibrated only for \Wqqprime decays and not for $\PZ(\PQq\PAQq)$ decays in data.
Therefore, additional uncertainties are applied when \ZH signal events enter the \PQb{}-tag \WH or \PW SRs.
Because the effect of these uncertainties on our results is small, we conservatively chose the size of the uncertainty to correspond to half the \PW tagger efficiency differences between \Wqqprime and $\PZ(\PQq\PAQq)$ decays in simulation.
The corrections to the jet tagging rates in the \PQb{}-veto SR and CRs derived from the VRs described in Section~\ref{sec:bkg_est_bveto}
are treated as systematic uncertainties, and are referred to as the ``nonclosure'' uncertainties.
The uncertainty in the top and diboson production cross sections~\cite{CMS:2021vhb,ATLAS:2018orx,CMS:2018amb,CMS:2019efc,CMS:2020gtj,CMS:2020mxy} produces about a 4\% effect on the \PQb{}-veto $\mathcal{R}_\mathrm{i}$.
Additional uncertainties associated with the choice of the renormalization and factorization scales \muR and \muF
were assessed by varying them independently up and down by a factor of 2, ignoring the case in which one parameter is scaled up while the other is scaled down.
Several other sources of uncertainties related to jet energy scale and resolution, \ptmiss modeling, effects of pileup, and choice of PDF were studied,
which collectively have less than 2\% impact on our predictions.
The uncertainty related to the choice of PDF is evaluated by reweighting the simulation using the variations in the corresponding NNPDF sets (3.0 or 3.1)~\cite{Ball:2014uwa,Ball:2017nwa}.
The statistical treatment of the systematic uncertainties is described in Section~\ref{sec:results}.
Summaries of the dominant systematic uncertainties and their impacts on the yields of the various sources of background and the signal are presented in Table~\ref{tab:syst_summary}.

\begin{table*}[ht!]
\centering
\topcaption{
The dominant systematic uncertainties and their effects on event yields (in \%) in various SRs.
For the 0- and \oneres backgrounds in the \PQb{}-veto SR, uncertainties are presented separately depending on the CR region used for the estimation.
A dash (\NA) indicates that the source of uncertainty is not applicable.
}
\label{tab:syst_summary}
\begin{tabular}{rrrrrcrrrr} 
\multirow{3}{*}{Source}   & \multicolumn{4}{c}{\PQb veto}                             & & \multicolumn{4}{c}{\PQb tag}            \\
                            \cline{2-5}                                                   \cline{7-10}
						  & \multicolumn{2}{c}{0- and \oneres bkg.} & Rare   & Signal & & Top       & \zerores  & Rare   & Signal \\
                          & 0-tag CR & 1-tag CR                     &        &        & & quark     &           &        &        \\ \hline
Integr.\ luminosity       & \NA      & \NA                          & 1.6    & 1.6    & & \NA       & \NA       & 1.6    & 1.6    \\
CR data size              & 6--71    & 5--50                        & \NA    & \NA    & & 3--100    & 2--35     & \NA    & \NA    \\
MC sample size            & 8--25    & 8--30                        & 14--24 & 2--5   & & 2--28     & 3--40     & 4--27  & 2--5   \\
\muR and \muF             & 1.2      & 0.4                          & 8      & ${<}5$ & & 2--10     & 0.5       & 11     & ${<}5$ \\
Trigger efficiency        & \NA      & \NA                          & 2--3   & 2--3   & & \NA       & \NA       & 2--3   & 2--3   \\
\PQb{}-tag correction     & ${<}1$   & ${<}1$                       & ${<}1$ & 1      & & 1         & ${<}3$    & ${<}3$ & 2--3   \\
\bbbar-tag correction     & \NA      & \NA                          & \NA    & \NA    & & \NA       & 4         & 2--7   & 4      \\
\PW{}-tag correction      & 12--28   & 6--22                        & 11--15 & 15     & & 1         & 9         & 7      & 9      \\
\PV{}-tag correction      & 7--15    & 2--10                        & 1--4   & 2      & & \NA       & \NA       & \NA    & \NA    \\
\PW{}-tag nonclosure      & 3--48    & 3--48                        & \NA    & \NA    & & \NA       & \NA       & \NA    & \NA    \\
\PV{}-tag nonclosure      & 1--27    & \NA                          & \NA    & \NA    & & \NA       & \NA       & \NA    & \NA    \\
1-res cross sections      & 2        & 4                            & \NA    & \NA    & & \NA       & \NA       & \NA    & \NA    \\
Fast simulation           & \NA      & \NA                          & \NA    & 5      & & \NA       & \NA       & \NA    & 8      \\
\end{tabular}
\end{table*}

Among these uncertainties, except for those from the sizes of the CR data and MC samples, the two leading systematic uncertainties that affect the 95\% confidence level (\CL) upper limits on the signal production cross sections discussed in the next section
are the \PW{}-tag nonclosure and \PW{}-tag correction uncertainties.
These degrade the cross section upper limits by up to about 15 and 10\%, respectively, depending on the signal model and the SUSY particle masses.
The others have much smaller impacts.

\section{Results}
\label{sec:results}

\begin{figure*}[ht!]
	\centering
	\includegraphics[width=0.49\linewidth]{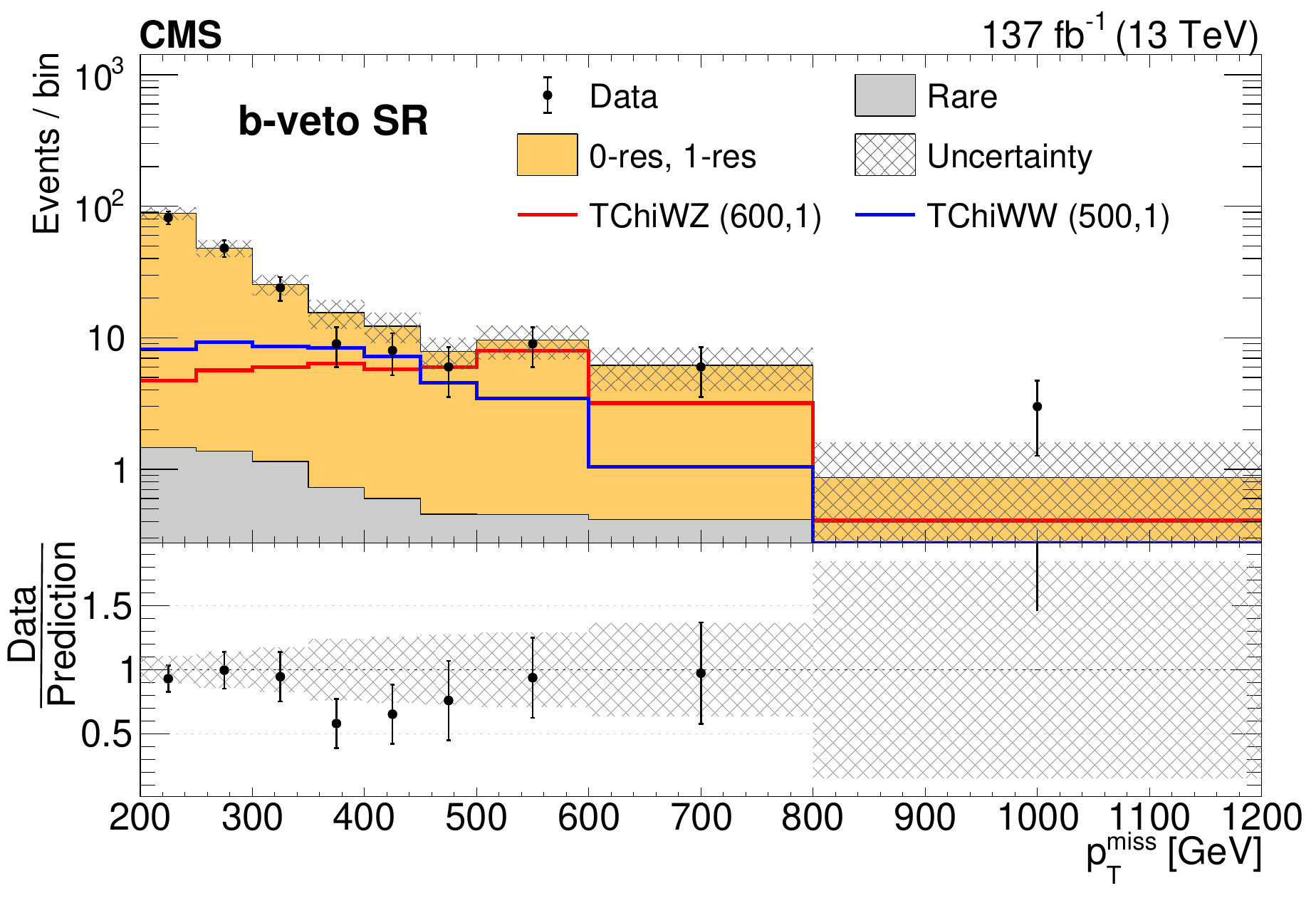}
	\includegraphics[width=0.49\linewidth]{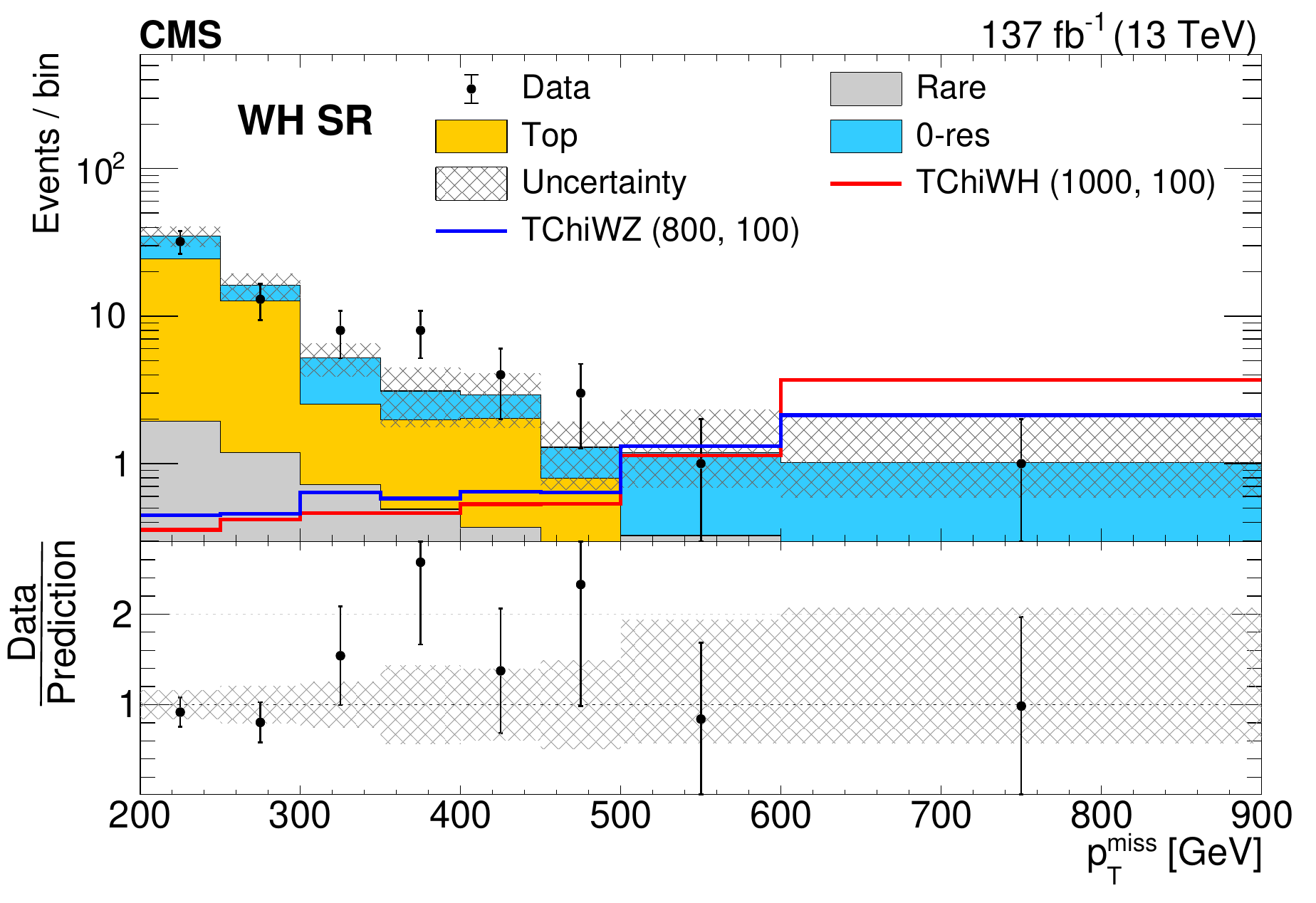}
	\includegraphics[width=0.49\linewidth]{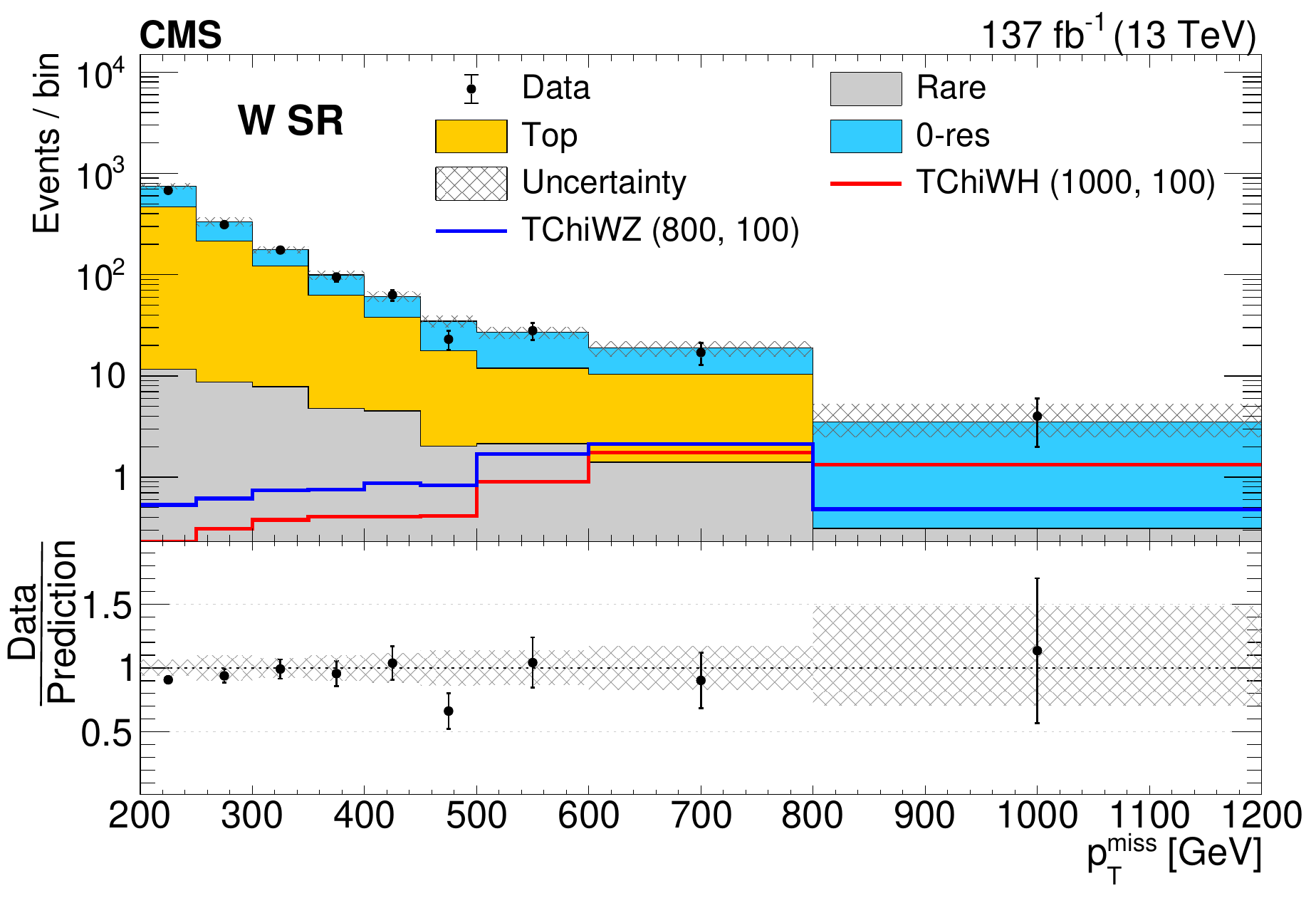}
	\includegraphics[width=0.49\linewidth]{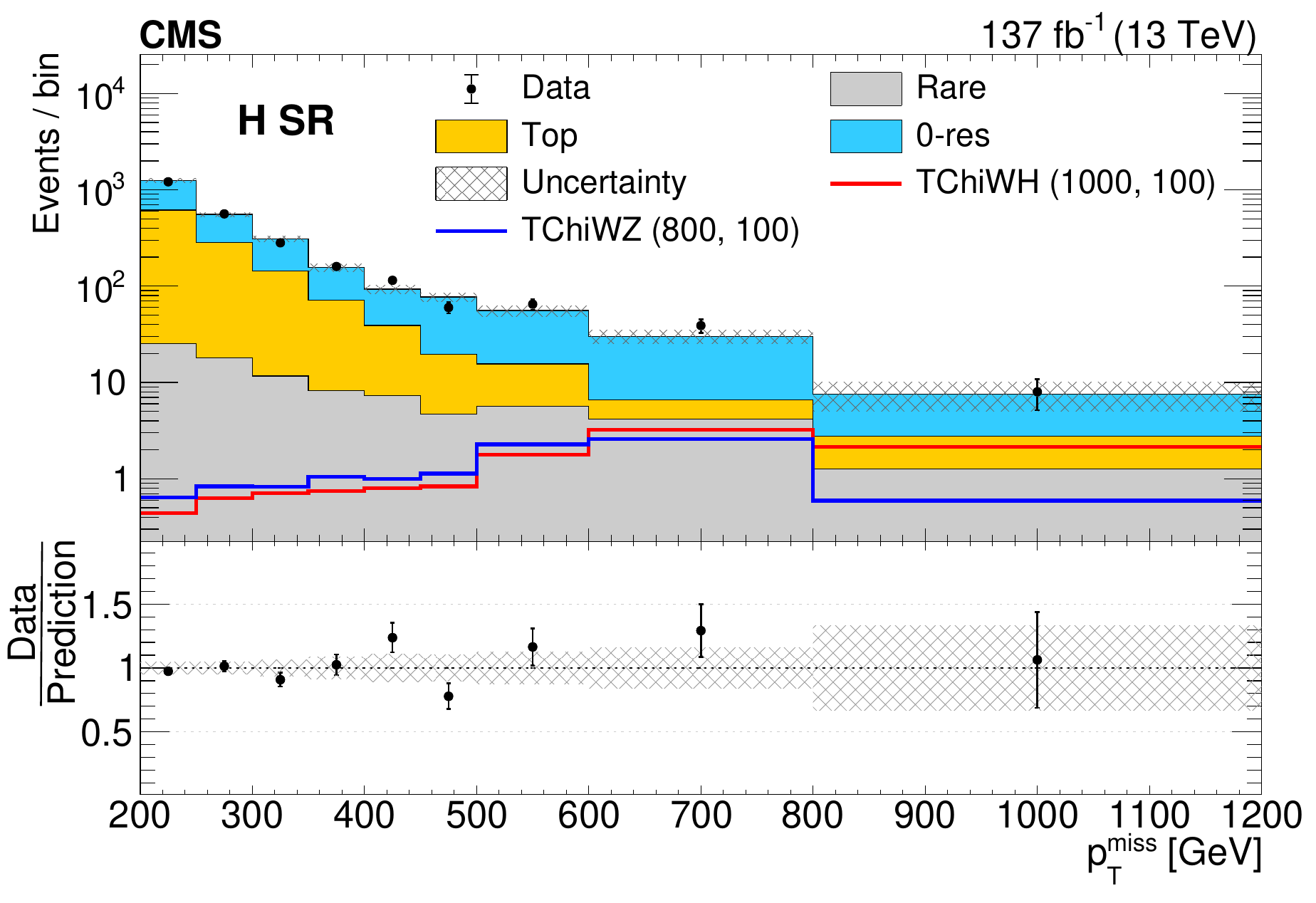}
	\caption{SM background prediction vs.\ observation in the \PQb{}-veto SR (upper left), the \WH SR (upper right), the \PW SR (lower left),
		and the \PH SR (lower right).
		The filled stacked histograms show the SM background predictions from the fit to the data in the CRs under the background-only hypothesis.
                The superimposed open histograms show the expectations for selected signal models, which are
		denoted in the legend by the name of the model followed by the assumed masses
		of the NLSP and LSP in \GeVns.
		The observed event yields are indicated by black markers.
		The hatched gray band corresponds to the total uncertainty in the prediction.
		}
\label{fig:predictions}
\end{figure*}

We perform simultaneous fits using a statistical model of our signal and SM background predictions.
The likelihood used for the fits is a product of Poisson distributions, one for each \ptmiss bin of each SR and CR.
The systematic uncertainties are included in the likelihood as nuisance parameters with log-normal constraints~\cite{Conway:2011in}, with the exception of the systematic uncertainties related to the finite size of the simulation, which use the method described in Ref.~\cite{Barlow-Beeston}.
This fitting procedure further constrains the predictions and the uncertainties in the predictions.
The yields of the SM backgrounds, determined from the fit applied only to the CRs under the background-only hypothesis, are shown along with the predicted yields of the signals and the observations in Fig.~\ref{fig:predictions}.
No statistically significant excess of events is observed in the data with respect to the SM background predictions.

We place 95\% \CL upper limits on the production cross section of either pairs of charginos or a chargino and a neutralino together.
The limits are computed based on a binned likelihood fit to the data in all of the \PQb{}-veto and \PQb{}-tag SRs and CRs, which takes into account the predicted background and signal yields.
A test statistic is used in conjunction with the \CLs criterion~\cite{CLS1,CLS2} to set upper limits.
The test statistic is the profile likelihood ratio, modified for upper limits~\cite{CMS-NOTE-2011-005}.
We compute limits using the asymptotic approximation~\cite{Cowan:2010js}.

\begin{figure*}[htb!]
	\centering
	\includegraphics[width=0.48\linewidth]{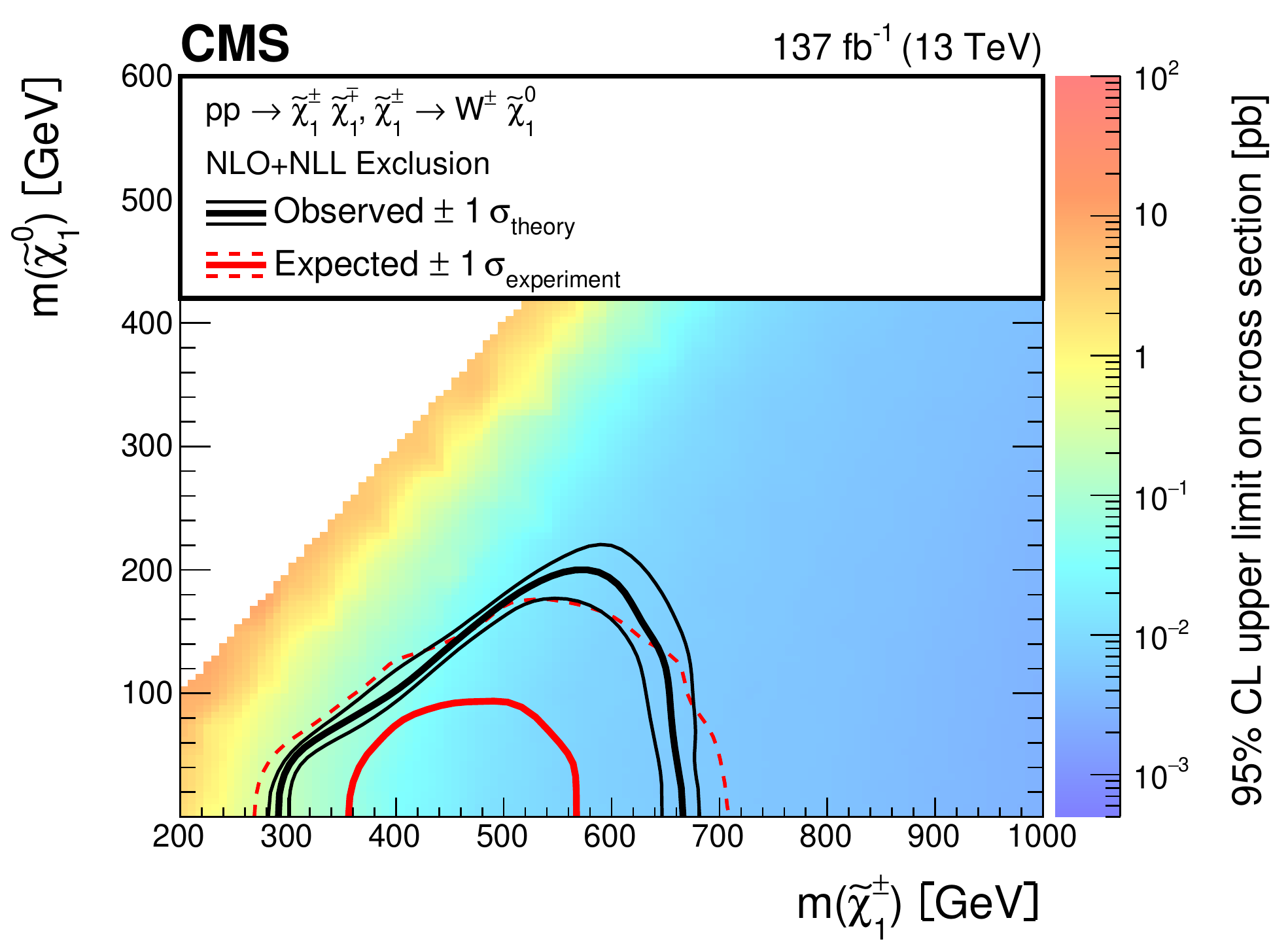}
	\includegraphics[width=0.48\linewidth]{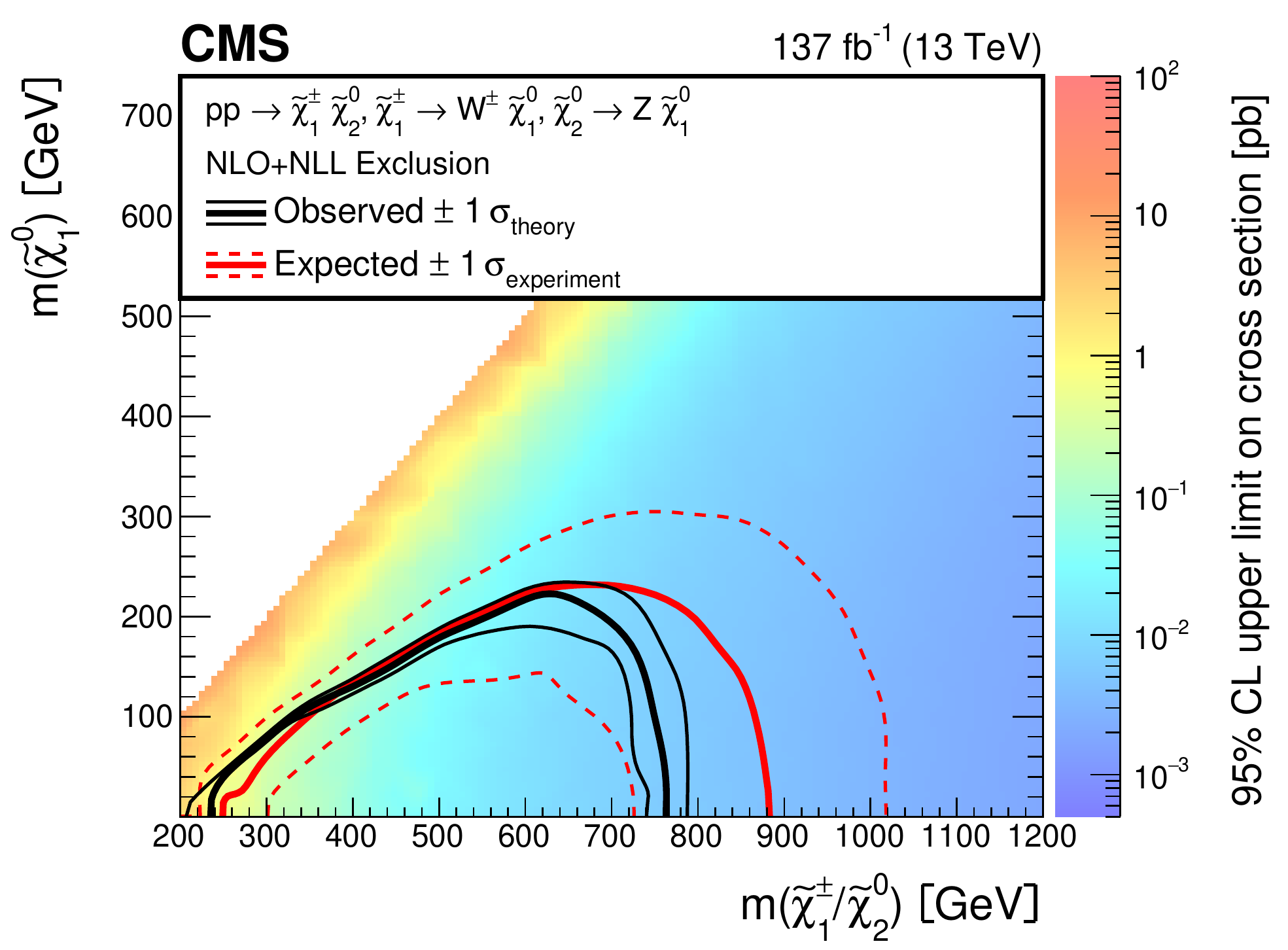}
	\includegraphics[width=0.48\linewidth]{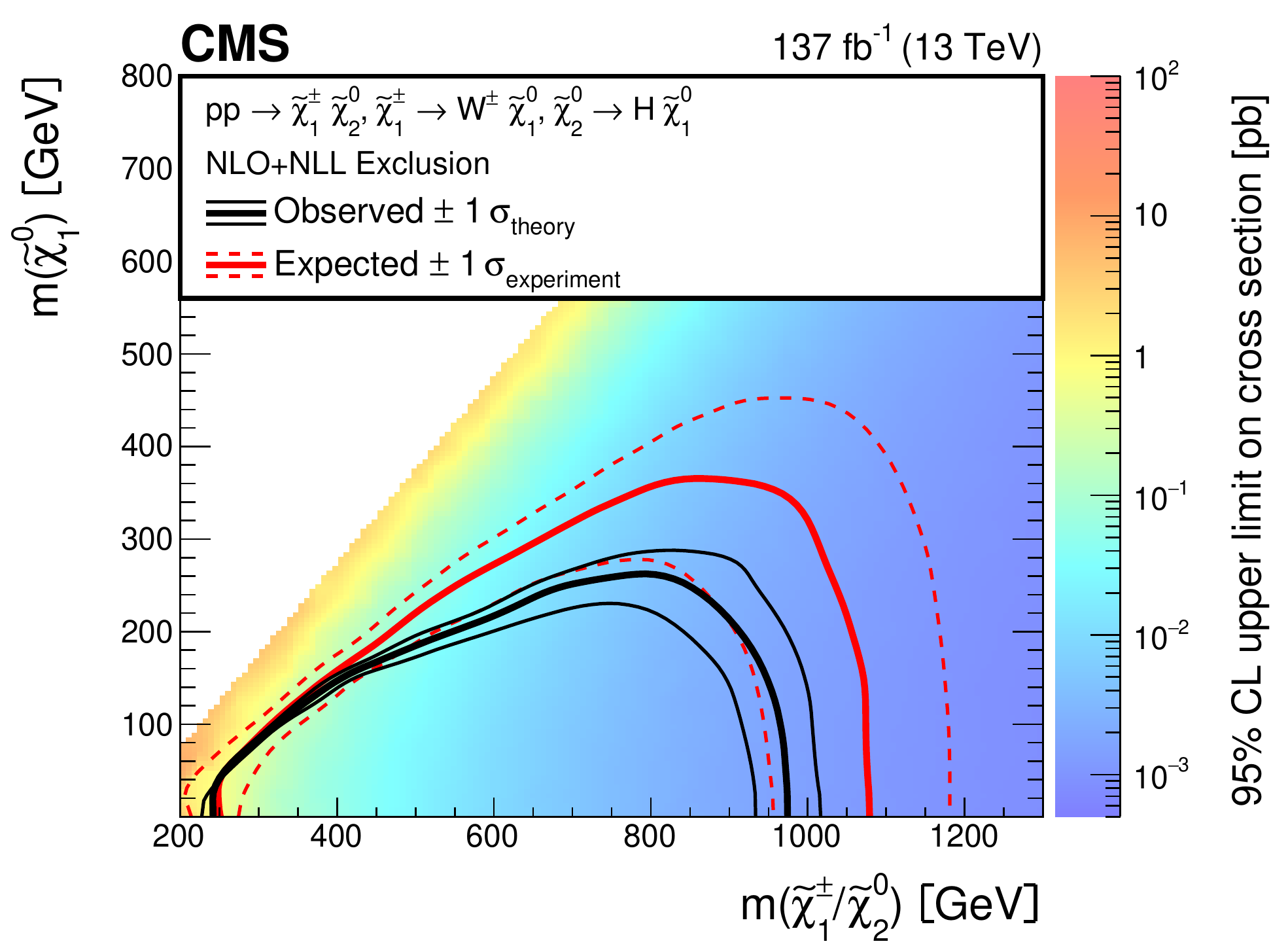}
	\caption{
                The 95\% \CL upper limits on
		the production cross sections for \firstchargpair assuming
		that each \firstcharg decays to a \PW boson and \firstchi (upper left, the TChiWW model) and \chargneut production assuming
		that the \firstcharg decays to a \PW boson and \firstchi and that the \secondchi decays to a \PZ boson and \firstchi (upper right, the TChiWZ model)
		or that the \secondchi decays to a Higgs boson and \firstchi (lower, the TChiWH model).
                The black curves represent the observed exclusion contour
                and the change in this contour due to variation of these
                cross sections within their theoretical uncertainties ($\sigma_{\text{theory}}$).
                The red curves indicate the mean expected exclusion contour
                and the region containing 68\% (${\pm}1\,\sigma_{\text{experiment}}$) of the expected
                exclusion limits under the background-only hypothesis.
		The mass exclusion limits are computed assuming wino-like cross sections.
		}
	\label{fig:limits1}
\end{figure*}

By comparing the upper limits on the production cross sections to the cross sections predicted for
chargino-pair production and for chargino-neutralino production, 95\% \CL mass exclusion contours
are derived.
The 95\% \CL mass exclusion contours within the NLSP-LSP mass plane are shown in Fig.~\ref{fig:limits1}
(upper left and upper right) for the TChiWW and TChiWZ models.
We exclude NLSP masses between 290 and 670\GeV assuming a pair of
charginos are produced and result in a pair of \PW bosons and a pair of light LSPs.
The observed limits are stronger than the expected limits because the observed event yields in the $300 < \ptmiss < 500\GeV$ region of the \PQb{}-veto SR are lower than the SM predictions.
For low-mass LSPs, we exclude NLSP masses between 230 and 760\GeV assuming chargino-neutralino production resulting in
\PW and \PZ bosons and a large \ptmiss.
For higher NLSP and lower LSP masses, the observed limits are weaker than the expected limits because the observed event yields in the last \ptmiss bin of the \PQb{}-veto SR are higher than the SM predictions.
In the case of chargino-neutralino production with a \PW boson and a Higgs boson, for low-mass LSPs we exclude NLSP
masses between 240 and 970\GeV.
Figure~\ref{fig:limits1} (lower) shows limits in the NLSP-LSP mass plane for the TChiWH model.
The observed limits are slightly weaker than the expected limits because a few intermediate \ptmiss bins of the \WH SR contain more observed events than expected.

\begin{figure}[htb!]
\centering
\includegraphics[width=0.48\textwidth]{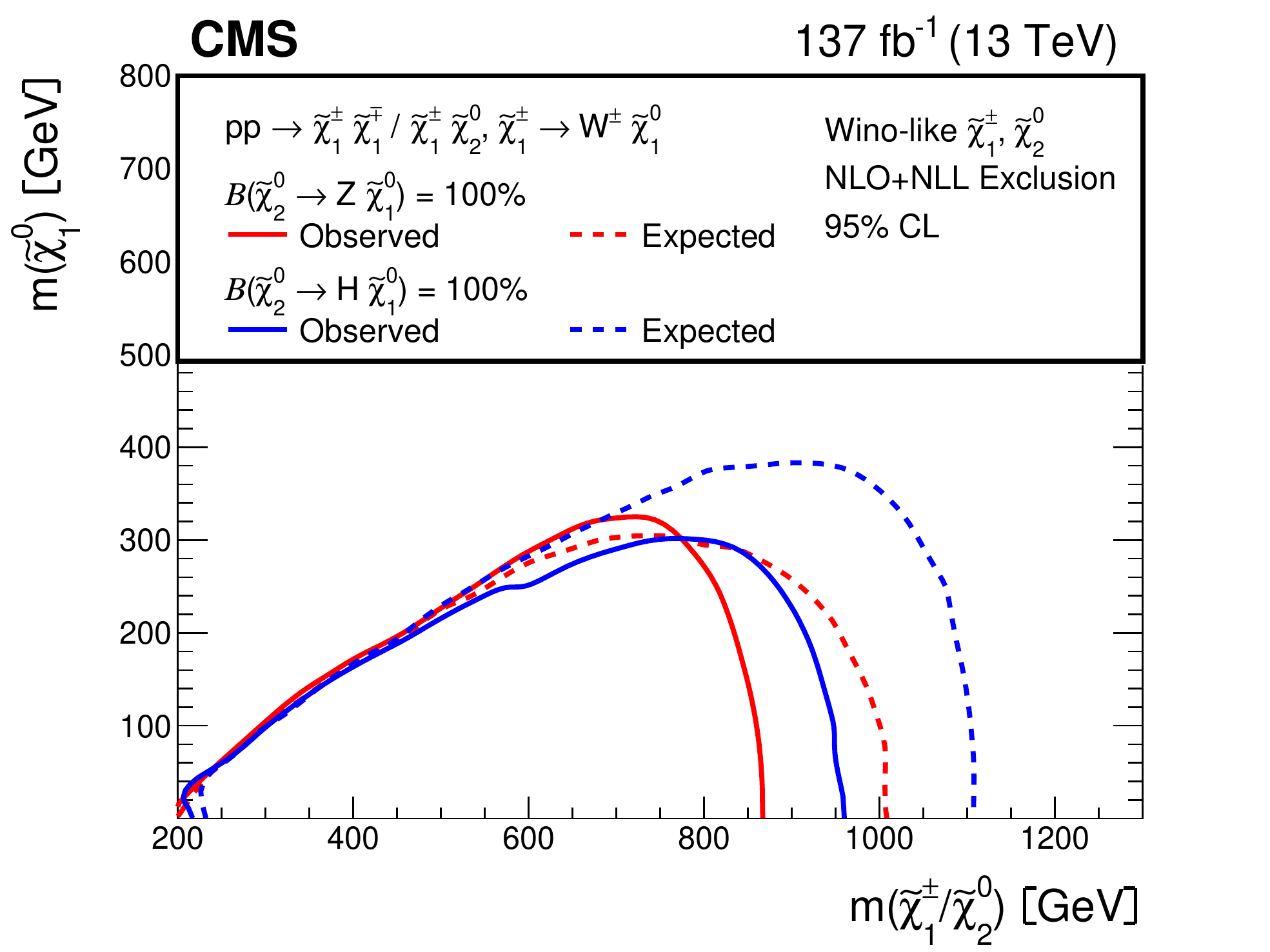}
\includegraphics[width=0.48\textwidth]{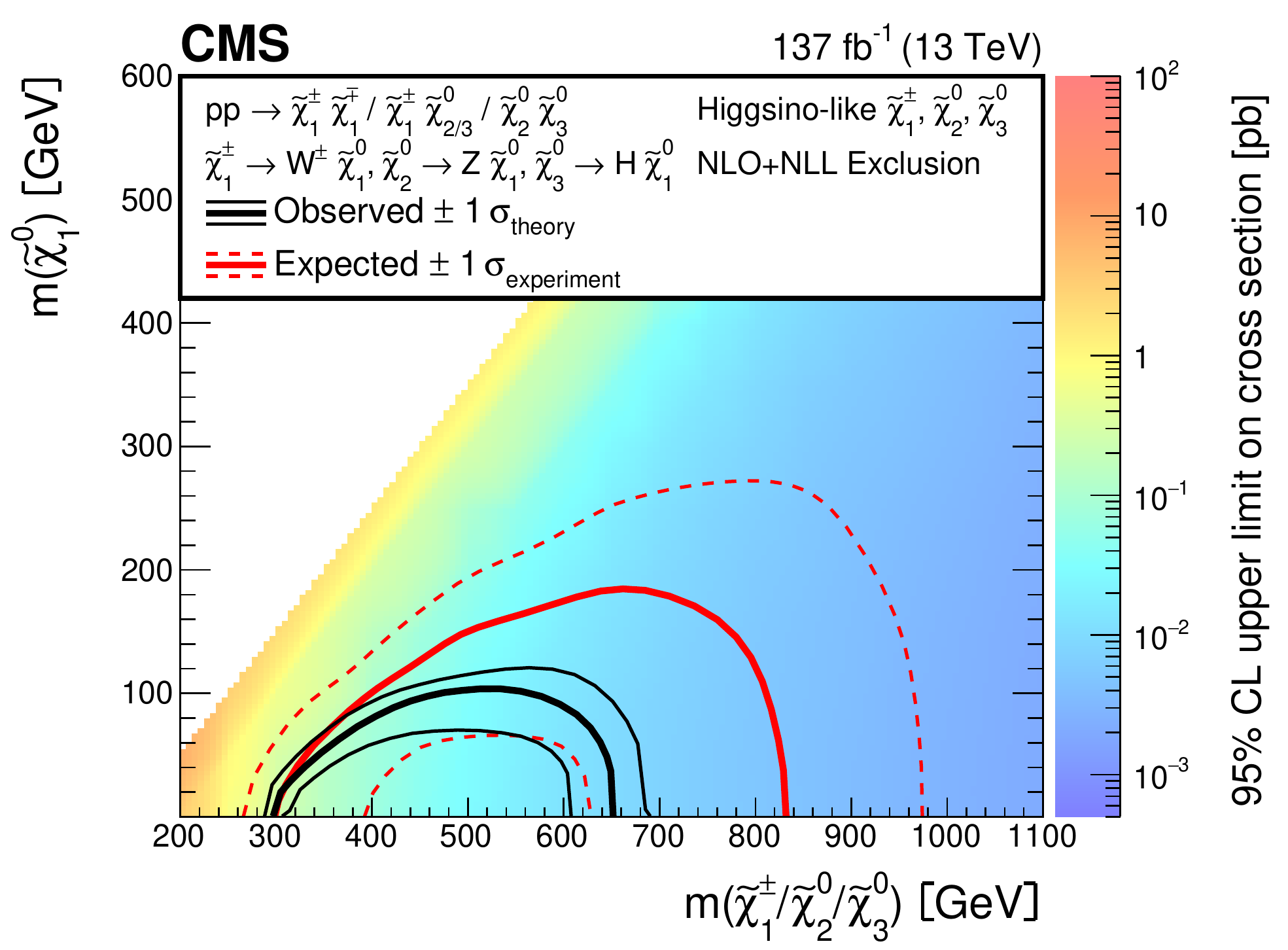}
\caption{Expected and observed 95\% \CL exclusion for mass-degenerate wino-like \firstchargpair and \chargneut
production (\cmsLeft) and higgsino-like \firstchargpair, \chargneut, \chargneutthree, and \neuttwoneutthree production
(\cmsRight) as functions of the NLSP and LSP masses.
The \firstcharg, \secondchi, and \thirdchi are considered to be mass degenerate.
For the higgsino-like case (\cmsRight), the 95\% \CL upper limits on the production cross sections are also shown,
but they are not shown for the wino-like case (\cmsLeft) because there are two distinct sets of limits depending on the chargino decay mode.
}
\label{fig:limits3}
\end{figure}

We also consider models including both \firstchargpair and \chargneut production
where the \firstcharg and \secondchi are the mass-degenerate wino-like NLSPs.
The expected and observed mass exclusions are presented in Fig.~\ref{fig:limits3} (\cmsLeft).
For the scenarios of $\secondchi\to\PZ\firstchi$ or $\secondchi\to\PH\firstchi$,
wino-like NLSP masses up to 870 and 960\GeV are excluded, respectively,
while exclusions up to 1010 and 1110\GeV are expected under the background-only hypothesis.

Figure~\ref{fig:limits3} (\cmsRight) shows the results for the mass-degenerate higgsino-like
NLSP scenario including
\firstchargpair, \chargneut, \chargneutthree, and \neuttwoneutthree production.
In this scenario, higgsino-like NLSP masses from 300 to 650\GeV are excluded.
The expected exclusion
reaches from 320 to 810\GeV. The observed exclusion is weaker than expected mainly because of a modest excess in the data over the background prediction in
the $300<\ptmiss<500\GeV$ region of the \WH SR.

\section{Summary}

A search is presented for signatures of electroweak production of charginos and neutralinos in fully hadronic final states.
The charginos are assumed to decay to the \PW boson and the lightest neutralino \firstchi,
and the heavier neutralinos (\secondchi and \thirdchi) are assumed to decay to either the \PZ or Higgs boson (\PH) and \firstchi.
The decay products of \PW, \PZ, or Higgs bosons are clustered into large-radius jets.
These jets are categorized based on their mass and a collection of novel jet-tagging algorithms based on deep neural networks.
Four search regions, three that require \PQb tags and one that excludes \PQb tags, are constructed to look for chargino- and neutralino-mediated production of
a pair of bosons, \WW, \WZ, or \WH,
together with a large transverse momentum imbalance.
We consider simplified models in which the charginos \firstcharg and the next-to-lightest
neutralino \secondchi are assumed to be the mass-degenerate next-to-lightest supersymmetric particles (NLSPs).
The lightest neutralino \firstchi is assumed to be bino-like and to be the lightest supersymmetric particle (LSP).
No statistically significant excess of events is observed in the data with respect to the expectation from the standard model.

Using wino-like pair production cross sections, 95\% confidence level mass exclusions are derived.
For signals with \WW, \WZ, or \WH boson pairs,
the NLSP mass exclusion limit for low-mass LSPs extends up to 670, 760, and 970\GeV, respectively.
When we consider models including both wino-like NLSP \chargneut and \firstchargpair production
under the assumption that either $\secondchi\to\PZ\firstchi$ or $\secondchi\to\PH\firstchi$,
the NLSP mass exclusion extends up to 870 and 960\GeV, respectively.
Alternatively, with higgsino-like NLSPs \firstcharg, \secondchi, and \thirdchi,
the higgsino masses from 300 to 650\GeV are excluded for low-mass LSPs.
These mass exclusions significantly improve on those achieved by searches using leptonic probes of SUSY for high NLSP masses.

\begin{acknowledgments}
We congratulate our colleagues in the CERN accelerator departments for the excellent performance of the LHC and thank the technical and administrative staffs at CERN and at other CMS institutes for their contributions to the success of the CMS effort. In addition, we gratefully acknowledge the computing centers and personnel of the Worldwide LHC Computing Grid and other centers for delivering so effectively the computing infrastructure essential to our analyses. Finally, we acknowledge the enduring support for the construction and operation of the LHC, the CMS detector, and the supporting computing infrastructure provided by the following funding agencies: BMBWF and FWF (Austria); FNRS and FWO (Belgium); CNPq, CAPES, FAPERJ, FAPERGS, and FAPESP (Brazil); MES and BNSF (Bulgaria); CERN; CAS, MoST, and NSFC (China); MINCIENCIAS (Colombia); MSES and CSF (Croatia); RIF (Cyprus); SENESCYT (Ecuador); MoER, ERC PUT and ERDF (Estonia); Academy of Finland, MEC, and HIP (Finland); CEA and CNRS/IN2P3 (France); BMBF, DFG, and HGF (Germany); GSRI (Greece); NKFIH (Hungary); DAE and DST (India); IPM (Iran); SFI (Ireland); INFN (Italy); MSIP and NRF (Republic of Korea); MES (Latvia); LAS (Lithuania); MOE and UM (Malaysia); BUAP, CINVESTAV, CONACYT, LNS, SEP, and UASLP-FAI (Mexico); MOS (Montenegro); MBIE (New Zealand); PAEC (Pakistan); MES and NSC (Poland); FCT (Portugal); MESTD (Serbia); MCIN/AEI and PCTI (Spain); MOSTR (Sri Lanka); Swiss Funding Agencies (Switzerland); MST (Taipei); MHESI and NSTDA (Thailand); TUBITAK and TENMAK (Turkey); NASU (Ukraine); STFC (United Kingdom); DOE and NSF (USA).

\hyphenation{Rachada-pisek} Individuals have received support from the Marie-Curie program and the European Research Council and Horizon 2020 Grant, contract Nos.\ 675440, 724704, 752730, 758316, 765710, 824093, 884104, and COST Action CA16108 (European Union); the Leventis Foundation; the Alfred P.\ Sloan Foundation; the Alexander von Humboldt Foundation; the Belgian Federal Science Policy Office; the Fonds pour la Formation \`a la Recherche dans l'Industrie et dans l'Agriculture (FRIA-Belgium); the Agentschap voor Innovatie door Wetenschap en Technologie (IWT-Belgium); the F.R.S.-FNRS and FWO (Belgium) under the ``Excellence of Science -- EOS" -- be.h project n.\ 30820817; the Beijing Municipal Science \& Technology Commission, No. Z191100007219010; the Ministry of Education, Youth and Sports (MEYS) of the Czech Republic; the Hellenic Foundation for Research and Innovation (HFRI), Project Number 2288 (Greece); the Deutsche Forschungsgemeinschaft (DFG), under Germany's Excellence Strategy -- EXC 2121 ``Quantum Universe" -- 390833306, and under project number 400140256 - GRK2497; the Hungarian Academy of Sciences, the New National Excellence Program - \'UNKP, the NKFIH research grants K 124845, K 124850, K 128713, K 128786, K 129058, K 131991, K 133046, K 138136, K 143460, K 143477, 2020-2.2.1-ED-2021-00181, and TKP2021-NKTA-64 (Hungary); the Council of Science and Industrial Research, India; the Latvian Council of Science; the Ministry of Education and Science, project no. 2022/WK/14, and the National Science Center, contracts Opus 2021/41/B/ST2/01369 and 2021/43/B/ST2/01552 (Poland); the Funda\c{c}\~ao para a Ci\^encia e a Tecnologia, grant CEECIND/01334/2018 (Portugal); the National Priorities Research Program by Qatar National Research Fund; MCIN/AEI/10.13039/501100011033, ERDF ``a way of making Europe", and the Programa Estatal de Fomento de la Investigaci{\'o}n Cient{\'i}fica y T{\'e}cnica de Excelencia Mar\'{\i}a de Maeztu, grant MDM-2017-0765 and Programa Severo Ochoa del Principado de Asturias (Spain); the Chulalongkorn Academic into Its 2nd Century Project Advancement Project, and the National Science, Research and Innovation Fund via the Program Management Unit for Human Resources \& Institutional Development, Research and Innovation, grant B05F650021 (Thailand); the Kavli Foundation; the Nvidia Corporation; the SuperMicro Corporation; the Welch Foundation, contract C-1845; and the Weston Havens Foundation (USA).
\end{acknowledgments}

\bibliography{auto_generated}
\cleardoublepage \appendix\section{The CMS Collaboration \label{app:collab}}\begin{sloppypar}\hyphenpenalty=5000\widowpenalty=500\clubpenalty=5000
\cmsinstitute{Yerevan Physics Institute, Yerevan, Armenia}
{\tolerance=6000
A.~Tumasyan\cmsAuthorMark{1}\cmsorcid{0009-0000-0684-6742}
\par}
\cmsinstitute{Institut f\"{u}r Hochenergiephysik, Vienna, Austria}
{\tolerance=6000
W.~Adam\cmsorcid{0000-0001-9099-4341}, J.W.~Andrejkovic, T.~Bergauer\cmsorcid{0000-0002-5786-0293}, S.~Chatterjee\cmsorcid{0000-0003-2660-0349}, K.~Damanakis\cmsorcid{0000-0001-5389-2872}, M.~Dragicevic\cmsorcid{0000-0003-1967-6783}, A.~Escalante~Del~Valle\cmsorcid{0000-0002-9702-6359}, R.~Fr\"{u}hwirth\cmsAuthorMark{2}\cmsorcid{0000-0002-0054-3369}, M.~Jeitler\cmsAuthorMark{2}\cmsorcid{0000-0002-5141-9560}, N.~Krammer\cmsorcid{0000-0002-0548-0985}, L.~Lechner\cmsorcid{0000-0002-3065-1141}, D.~Liko\cmsorcid{0000-0002-3380-473X}, I.~Mikulec\cmsorcid{0000-0003-0385-2746}, P.~Paulitsch, F.M.~Pitters, J.~Schieck\cmsAuthorMark{2}\cmsorcid{0000-0002-1058-8093}, R.~Sch\"{o}fbeck\cmsorcid{0000-0002-2332-8784}, D.~Schwarz\cmsorcid{0000-0002-3821-7331}, S.~Templ\cmsorcid{0000-0003-3137-5692}, W.~Waltenberger\cmsorcid{0000-0002-6215-7228}, C.-E.~Wulz\cmsAuthorMark{2}\cmsorcid{0000-0001-9226-5812}
% ...
\par}
\vskip\cmsinstskip
\dag:~Deceased\\
$^{1}$Also at Yerevan State University, Yerevan, Armenia\\
$^{2}$Also at TU Wien, Vienna, Austria\\
% ...
\end{sloppypar}
\end{document}